\documentclass[fleqn,usenatbib]{mnras}

\usepackage{newtxtext,newtxmath}
\usepackage{amsmath}

\usepackage[T1]{fontenc}

\DeclareRobustCommand{\VAN}[3]{#2}
\let\VANthebibliography\thebibliography
\def\thebibliography{\DeclareRobustCommand{\VAN}[3]{##3}\VANthebibliography}

\usepackage{graphicx}	
\usepackage{amsmath}	

\newcommand{\kepler}{\textit{Kepler}}
\newcommand{\ktwo}{\textit{K2}}
\newcommand{\gaia}{\textit{Gaia}}
\newcommand{\hst}{\textit{HST}}
\newcommand{\iue}{\textit{IUE}}
\newcommand{\galex}{\textit{GALEX}}

\newcommand{\tess}{\textit{TESS}}
\newcommand{\swift}{\textit{Swift}}
\newcommand{\angstrom}{\mbox{\normalfont\AA}}

\newcommand{\water}{$\mathrm{H_{2}O}$}
\newcommand{\carbondioxide}{$\mathrm{CO_{2}}$}
\newcommand{\oxygen}{$\mathrm{O_{2}}$}

\newcommand{\Msun}{$\mathrm{M_{\odot}}$}

\title[Testing flare models with \tess\ and \galex\ ]{Extending Optical Flare Models to the UV: Results from Comparing of \tess\ and \galex\ Flare Observations \textcolor{black}{for M Dwarfs}}

\author[J. A. G. Jackman et al.]{
James A. G. Jackman$^{1}$\thanks{E-mail: jamesjackman@asu.edu (JAGJ)},
Evgenya Shkolnik$^{1}$,
Chase Million $^{2}$,
Scott Fleming $^{3}$,\newauthor
Tyler Richey-Yowell$^{4,1}$\thanks{Percival Lowell Postdoctoral Fellow},
Parke Loyd$^{5}$\\
\\
$^{1}$School of Earth and Space Exploration, Arizona State University, Tempe, AZ, 85287, USA\\
$^{2}$ Million Concepts LLC, 1355 Bardstown Road, No. 132, Louisville, KY, 40204, USA \\
$^{3}$ Space Telescope Science Institute, 3700 San Martin Drive, Baltimore, MD, 21218, USA \\
$^{4}$ Lowell Observatory, 1400 W Mars Hill Rd, Flagstaff, AZ, 86001, USA\\
$^{5}$ Eureka Scientific, 2452 Delmer Street Suite 100, Oakland, CA, 94602-3017, USA \\
}

\date{Accepted XXX. Received YYY; in original form ZZZ}

\pubyear{2021}

\begin{document}
\label{firstpage}
\pagerange{\pageref{firstpage}--\pageref{lastpage}}
\maketitle

\begin{abstract}
The ultraviolet (UV) emission of stellar flares \textcolor{black}{may have a }
pivotal role in the habitability of rocky exoplanets around low-mass stars. Previous studies have used white-light observations to calibrate empirical models \textcolor{black}{which describe the optical and UV flare emission. }
However, the accuracy of the UV predictions of models have \textcolor{black}{previously} not been tested. 
We combined \tess\ optical and \galex\ UV observations to test the UV predictions of empirical flare models calibrated using optical flare rates of M stars. We find that the canonical 9000\,K blackbody model used by flare studies underestimates the \textcolor{black}{\galex} NUV energies of field age M stars by up to a factor of \textcolor{black}{$6.5\pm0.7$} and the \textcolor{black}{\galex} FUV energies of fully convective field age M stars by \textcolor{black}{$30.6\pm10.0$}. 
We calculated energy correction factors that can be used to bring the UV predictions of flare models closer in line with observations. We calculated pseudo-continuum flare temperatures that describe both the white-light and \textcolor{black}{\galex} NUV emission. We measured a temperature of 10,700\,K for flares from fully convective M stars after accounting for the contribution from UV line emission. We also applied our correction factors to the results of previous studies of the role of flares in abiogenesis. Our results show that M stars do not need to be as active as previously thought in order to provide the NUV flux required for prebiotic chemistry, however we note that flares will also provide more FUV flux than previously modelled.

\end{abstract}




\section{Introduction}
Stellar flares have become a topic of ardent research in recent years, in part due to their potential role in the habitability of exoplanets around low-mass stars. Flares are caused by magnetic reconnection events in the outer atmospheres of stars \citep[e.g.][]{Benz10}. The energy released in these events accelerates charged particles from the reconnection site down towards the chromosphere. These particles impact the dense chromospheric plasma at sites termed the flare footpoints, resulting in rapid heating and evaporation \citep[e.g.][]{Fisher85,Milligan06}. The evaporated plasma rises to fill the newly reconnected field lines, which are anchored into the chromosphere at the flare footpoints. At the same time a descending compression known as a chromospheric condensation is formed that pushes towards the lower chromosphere \citep[e.g.][]{Fisher85,Kowalski18}. The entire flare process releases energy from radio wavelengths up to hard X-rays and even gamma ray emission \citep[e.g.][]{Hurford03,Kumar16}. The white-light emission from Solar and stellar flares, termed ``white-light flares'', is believed to be associated with the flare footpoints \citep[][]{Fletcher07,Krucker15}. However, the exact mechanism for the white-light emission is still a matter of debate. 
The white-light emission may be due to direct heating of the photosphere via non-thermal electrons \citep[e.g.][]{Hudson72,Neidig89}. 
However, only the highest energy electrons are expected to be able to penetrate down to the lower chromosphere/photosphere. \textcolor{black}{Instead, lower energy electrons may trigger the white-light emission either directly through the heating of layers in the mid/upper chromosphere \citep[e.g.][]{Kerr14,Jurvak18}, or indirectly through these heated layers backheating lower layers of atmosphere, e.g. via descending chromospheric condensations \citep[e.g.][]{Kowalski18}.} 
Studies have also observed hard X-ray flares that appear to lack white-light counterparts, suggesting that the presence of white-light emission may depend on factors such as the local magnetic field strength, local plasma conditions and the rate of energy deposition \citep[][]{Watanabe17,Watanabe20}. 

Although the first modern detection of a Solar flare was in the optical \citep[][]{Carrington_1859}, the detection of \textcolor{black}{white-light} flares is hindered by the contrast between the white-light emission and the Solar photosphere. This can bias detections to off-limb or high energy events, where isolation of the white-light emission can be done without ambiguity \citep[e.g.][]{Zhao21}. 
However, dedicated studies have found evidence of white-light emission from both low and high energy Solar flares alike \citep[][]{Hudson06}. It is the white-light emission that we regularly detect and study in stellar flares from low-mass stars \citep[e.g.][]{Gunther20}. These stars can have white-light flares with bolometric energies equal to, or greater, than those seen from the modern Sun \citep[e.g. $10^{32}$ erg;][]{Carrington_1859,Carrington_Energy}. 
The cooler temperatures and lower photospheric luminosities of low-mass stars 
make flares appear larger in amplitude within a given photometric filter than they would for an equal energy Solar flare. Consequently, the white-light emission has become a tracer for studying flare activity on low-mass stars.

\textcolor{black}{Studies have used data from wide-field exoplanet surveys such as \kepler, NGTS, and \tess\  \citep[][]{Borucki10,Ricker14,Wheatley18} to investigate the energies and rates of stellar flares. These surveys simultaneously observe tens to hundreds of thousands of stars for durations of weeks to months (e.g. \tess, NGTS) or years (e.g. \kepler), enabling the detection of large samples of stellar flares. Previous works have used these datasets to investigate how flare properties such as amplitude, duration and energy change 
across spectral types \citep[e.g.][]{Yang17,Jackman21}, and how flare rates change with age, 
finding that younger stars flare more often than their older counterparts \citep[e.g.][]{Davenport19,Ilin19,Feinstein20}.  }  

The white-light emission of flares is often approximated \textcolor{black}{in works using single-bandpass photometry} 
as a blackbody with a continuum temperature of 9000\,K \citep[e.g.][]{Hawley92,Shibayama13,Gunther20}. 
While this continuum emission dominates the optical energy budget (\textcolor{black}{particularly above above $\approx$4000\AA}), white-light flares also release energy via emission lines from a variety of species such as He, Na and Fe \citep[e.g.][]{Fuhrmeister11,Muheki20}. 
However, the most notable line feature is the Balmer series, in particular the Balmer jump at $\approx$ 3650\AA. 
At the Balmer jump the flare \textcolor{black}{flux can} increase above the \textcolor{black}{level predicted by blackbody models fitted to the blue-optical continuum \citep[e.g.][]{Kowalski13,Kowalski19}.} 
The elevated continuum emission persists into the near-ultraviolet (NUV ;$\approx$2000-3000\AA),
suggesting that the white-light and NUV emission arise from the same heated atmospheric layers \citep[e.g.][]{Joshi21}. The NUV emission of flares includes a greater contribution from emission lines than in the optical, notably from Mg II and Fe II. \citet{Hawley07} found that these lines contributed broadly equal levels of flux. \textcolor{black}{They measured that emission lines contributed between 20 and 50 per cent of the NUV flux, }
with higher energy flares being more continuum-dominated. \citet{Kowalski19} measured line contributions of approximately 40 per cent in the 2510-2841\AA\ range from two flares from GJ 1243 observed with \hst\ COS.

In the far-ultraviolet (FUV; 1150-1700\AA), 
studies have observed both strong continuum and line emission. Observations of M-star flares have shown that the FUV can precede the white-light emission, suggesting it may arise from the initial heating, compression and evaporation of plasma at the flare footpoints \citep[][]{Hawley03,Froning19,MacGregor21}. In addition, studies of flares from M dwarfs have measured temperatures up to 20,000\,K and even 40,000\,K in both the optical and FUV \citep[e.g.][]{Loyd18,Froning19,Howard20}.

Studies of flares from low-mass stars have sought to understand \textcolor{black}{the} effects \textcolor{black}{of their UV emission} on the habitability of terrestrial exoplanets. 
\textcolor{black}{The NUV emission from flares may help drive prebiotic photochemistry on the surfaces of rocky exoplanets around low-mass stars, in particular for the formation of amino acids and RNA \citep[e.g.][]{Ranjan17, Rimmer18}. FUV photons can dissociate atmospheric molecules such as \water, \carbondioxide, and \oxygen, species commonly used as biosignatures in studies  characterising exoplanetary atmospheres \citep[e.g.][]{Hu12,Tian14}. FUV flare emission may also be responsible for breakdown of anoxic biosignatures such as prebiotic HCN, while repeated flare events may permanently alter atmospheric compositions \citep[][]{Venot16,Rimmer19}. These effects will complicate searches for biosignatures with telescopes such as \textit{JWST} and the Extremely Large Telescope \citep[e.g.][]{ Rugheimer15,Gialluca21}.} 

Measuring the ultraviolet (UV) flaring activity and rates of 
individual stars currently requires expensive campaigns with space-based telescopes such as \hst\ (NUV, FUV) and \swift\ (NUV only), limiting large scale surveys. 
Habitability studies have aimed to get around this by extrapolating white-light flare rates into the UV, to estimate UV flare rates that can then be put into photo-chemical models, atmospheric studies \citep[e.g.][]{Chen21}, or compared to empirical limits for prebiotic chemistry \citep[e.g.][]{Rimmer18,Gunther20,Murray22}. To calculate white-light flare rates, models of the plasma emission at the flare footprints inform renormalisation of the optical flare lightcurves \citep[e.g.][]{Shibayama13}. Renormalised models are then extrapolated into the UV \citep[e.g.][]{Feinstein20, Glazier20}.

\textcolor{black}{The models used to evaluate the UV effects of flares} 
span a range of complexities. They range from blackbody-only models often used for calculating bolometric energies from flares detected with single bandpass photometry \citep[e.g. \textcolor{black}{a} 9000\,K \textcolor{black}{blackbody};][]{Shibayama13} to those that combine these blackbody models with measurements from archival UV spectra \citep[e.g.][]{Loyd18muscles}. \textcolor{black}{However, the UV predictions of these models have not been well tested.} 
\citet{Kowalski19} found that the 9000\,K blackbody underestimated the NUV continuum and the total NUV emission in \hst\ COS observations of two flares by factors of 2 and 3 respectively. 
Along with this, models do not account for continuum temperatures that go above 9000\,K during the peaks of flares \citep[e.g.][]{Kowalski13,Howard20}, \textcolor{black}{and increase the UV flare flux. These results highlight the need for testing the UV predictions of current flare models.}

One \textcolor{black}{way to test} the UV predictions of empirical flare models calibrated using \textcolor{black}{optical} observations is to use archival data from \galex. \citet{Million16} presented the gPhoton python package, that enables users to create UV lightcurves from archival \galex\ data, \textcolor{black}{something previously limited to special request \citep[e.g.][]{Robinson05, Welsh07}.} \citet{Brasseur19} used these data to study the NUV flare properties from F to M type stars previously observed with \kepler, and measured the average NUV flare rate of stars in thair sample. 
If both the average white-light and UV flare rates can be measured for a group of stars, this can provide a way of testing the UV predictions of empirical flare models used by white-light flare studies. 

In this work we present the results of testing the UV predictions of \textcolor{black}{six} empirical flare models from the literature, using \tess\ white-light and \galex\ NUV and FUV observations. We used \tess\ optical observations to calibrate each flare model and predict the average UV flare rates of partially and fully convective M stars that were observed with both \tess\ and \galex, providing an opportunity to constrain the accuracy of these models in the UV. We will discuss the methods used to obtain the data, detrend lightcurves and detect flares. 
We describe the models we tested and their use in existing flare studies. We will then detail how we have tested the accuracy of each model, the results of our tests and the impact of these results on flare models and existing tests of exoplanet habitability.

\section{Data}
\textcolor{black}{In this section we discuss the optical and UV observations we used in our analysis. We also describe how we constructed a sample of M stars that could be used to test the UV predictions of flare models.}

\subsection{\tess}
The \textit{Transiting Exoplanet Survey Satellite} \citep[\tess;][]{Ricker14} is a space-based wide-field survey designed to search for the transits of exoplanets in front of their host stars. \tess\ began observations for its primary mission in July 2018 and completed them in July 2020, observing each ecliptic hemisphere for one year. \textcolor{black}{The first extended mission for \tess\ lasted from August 2020 to August 2022 and reobserved the southern and northern hemispheres, along with observing a portion of the ecliptic plane.} 
\tess\ observes in a series of sectors, with each sector being observed for approximately 27 days. Each sector has a total field of view of 24$\times$96 square degrees, which is split into four regions, with each region being observed by one of four cameras. Each camera has a pixel scale of 21\arcsec\ per pixel. During the primary mission \tess\ observed with two cadences, a 30 minute long cadence mode for full frame images and a 2 minute short cadence mode for postage stamps. In the first extended mission the long cadence mode full frame image depth was changed to 10 minutes and a new 20 second fast cadence mode was added. 

We used the 2 minute cadence \tess\ lightcurves from sectors 1 to 32 in this work. 
We elected to use the 2 minute cadence data \textcolor{black}{for sectors from the extended mission} for consistency with the data from the primary mission. This consistency across multiple sectors is important during our analysis of the efficiency of our flare detection method in Sect.\,\ref{sec:tess_flare_occ_method}. 
Lightcurves are automatically generated for all 2 minute cadence \tess\ targets using the \tess\ Science Processing Operations Center pipeline \citep[SPOC;][]{spoc16} and made publicly available on MAST. Similar to the lightcurve data products available from the \kepler\ mission, the \tess\ data products consist of both Simple Aperture Photometry (SAP) and Pre-Search Data Conditioned (PDC\_SAP) data. 
We used the PDC\_SAP data in this work. The PDC\_SAP lightcurves are filtered to remove long term trends due to possible systematic effects, while keeping shorter period astrophysical signals such as transits, eclipses and flares. These lightcurves were corrected in the SPOC pipeline for dilution from other stars in and around the \tess\ aperture, as denoted by the CROWDSAP value in the \tess\ lightcurve header files.

\subsection{\galex}
The \textit{Galaxy Evolution Explorer} \citep[\galex;][]{Martin05,Morrissey05} was a mission to study the UV characteristics of galaxies.  
\galex\ operated from 2003 April to 2012 June and observed about two-thirds of the sky at UV wavelengths. \galex\ had two direct-imaging filters \textcolor{black}{which we call} the \textcolor{black}{\galex} NUV (1771-2831\angstrom) and FUV (1344-1786\angstrom) \citep[][]{Morrissey05}---capable observing simultaneously by means of a dichroic mirror, in addition to a slitless spectroscopic grism mode. \textcolor{black}{We note that the \galex\ NUV and FUV bandpasses do not cover the entirety of NUV and FUV wavelengths, but rather a subset of them.} 
The majority \textcolor{black}{of} observations were made simultaneously \textcolor{black}{in both bandpasses} until the FUV detector stopped operating in 2009. \textcolor{black}{For the remainder of the mission lifetime observations continued with only the NUV band.}
The micro-channel plate detectors of \galex\ produced time-tagged photon lists that can be used to generate lightcurves for study of stellar UV variability at sub-minute time resolutions \citep[][]{Robinson05, Welsh07}, although this was not a normal data type produced by the mission. The gPhoton project now makes it possible to generate calibrated light curves from the time-tagged photon lists on demand, with customizable photometric aperture and cadence \citep[][]{Million16}. The gPhoton python package\footnote{\url{https://gphoton.readthedocs.io/en/master/}} has been used to study the UV characteristics of stellar flares from individual sources \citep[][]{Million16} and large samples \citep[][]{Brasseur19}, allowing for detailed comparison with optical flare studies.

\subsection{Sample selection} \label{sec:sample}
\begin{figure}
    \centering
    \includegraphics[width=\columnwidth]{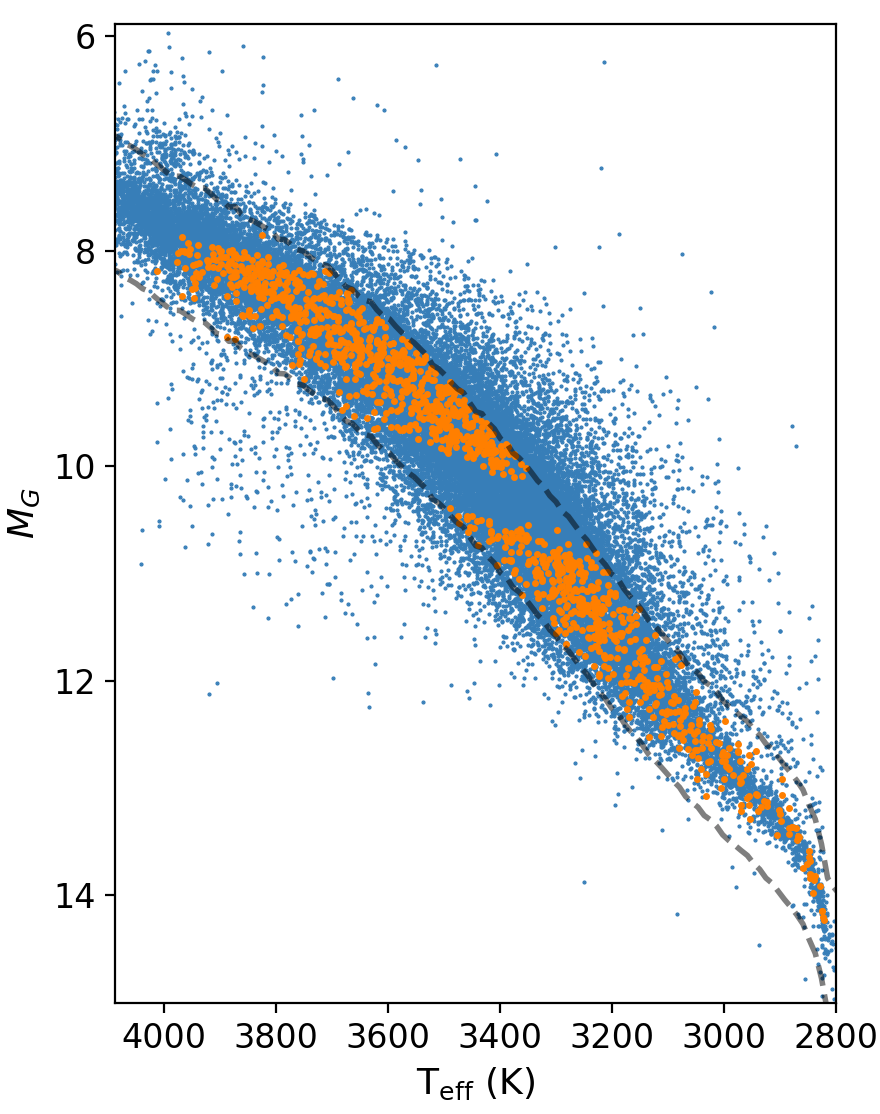}
    \caption{Hertzsprung-Russell diagram of the sample used in this work. The blue points are all stars that were observed with \tess\ in the 2-minute short cadence mode in sectors 1 to 32. The dashed grey lines are the upper and lower main-sequence limits outlined in Sect.\,\ref{sec:sample}. The orange points are the stars used in this work for testing flare models. The gap in the middle is where we avoided stars with masses spanning the fully convective boundary, where the stellar dynamo changes.}
    \label{fig:hr_diagram}
\end{figure}
In this work we focused on the flaring behaviour of main-sequence M stars. 
We obtained stellar properties for each star in our 2-min cadence sample from the \tess\ Input Catalogue (TIC) v8 \citep[][]{Stassun19}. \textcolor{black}{We then filtered our sample to remove all stars with listed masses above 0.6\Msun, along with those that didn't have listed mass, effective temperature or radius values.} 
To limit our sample to main sequence M stars we used a Hertzsprung-Russell (HR) diagram. To determine which stars were consistent with being an isolated main-sequence star, we first calculated a median absolute magnitude curve using all stars from the TIC v8 catalogue that resided within 100pc and passed the \gaia\ DR2 astrometric and photometric quality checks recommended by \citet{Arenou18} and \citet{Lindegren18}. This curve was measured as a function of effective temperature, taken from the TIC v8 catalogue. We then used the empirically defined main-sequence absolute magnitude limits from \citet{Jackman21} of $\Delta M_{G}=+0.65$ and $\Delta M_{G}=-0.55$ from the median curve to remove stars that were not consistent with being an isolated main-sequence M star. 
\textcolor{black}{This removed }equal-mass binary stars and young stars that have \textcolor{black}{not} contracted onto the main sequence \citep[e.g.][]{Baraffe18}. This resulted in a sample of \textcolor{black}{32347} stars.

To increase our chance of detecting 
flares in the \galex\ lightcurves, we followed the method of \citet{Brasseur19} and required that each star in our sample had at least 30 minutes of \galex\ NUV observations. We measured the \galex\ NUV exposure time of all 32347 stars using the gPhoton gFind tool and removed stars that were only observed with \galex\ for short durations \textcolor{black}{(less than 30 minutes in total)} or not at all. Stars observed for short durations would contribute only marginally to the sample and have a limited chance of a flare detection. 
This step resulted in \textcolor{black}{2189} M stars. We chose to split our sample into two mass ranges, using mass values from the TIC v8. These were 0.37-0.6\Msun\ and 0.1-0.29\Msun. These mass ranges correspond to M0-M2 and M4-M5 spectral types inclusive \citep[e.g.][]{Stassun19,Cifuentes20} and unambiguously sample either side of the fully convective boundary. The interior of M stars is thought to change from partially convective to fully convective at M3 \citep[0.3-0.35\Msun;][]{Chabrier97,Baraffe18,Macdonald18}. This change in the interior is accompanied by a change in the stellar dynamo \citep[e.g.][]{Shulyak15,Yadav15,Brown20}, which in turn may affect magnetic activity such as the flare occurrence rate \citep[e.g][]{Raetz20,Jackman21}. By choosing mass ranges below and above this transition region, we can separate our sample into partially and fully convective subsets without concern of contamination from a change in dynamo. 

Our filtering reduced our sample to 1250 stars. \textcolor{black}{Figure.\,\ref{fig:hr_diagram} shows the distribution of the filtered sample on the HR diagram.} 
758 \textcolor{black}{stars in our sample} have masses between 0.37 and \textcolor{black}{0.6} \Msun, and  \textcolor{black}{492} have masses between 0.1 and 0.29\Msun. 
We used these samples to study the relations between the optical and \galex\ NUV flare behaviour. Our samples for studying the \galex\ FUV behaviour comprise subsets of these, due to there being less \galex\ FUV data available. For stars with \galex\ FUV data, there were \textcolor{black}{484} stars with masses between 0.37 and 0.6 \Msun\ and \textcolor{black}{312} stars with masses between 0.1 and 0.29 \Msun.

For each target in our filtered sample, we downloaded the \tess\ PDC\_SAP short cadence lightcurves using the Python lightkurve tool. 
We generated 30-second cadence \galex\ lightcurves using the gAperture function in gPhoton with a standard aperture size of 12.8\arcsec and a background annulus with inner and outer radii of 25.6 and 51.2\arcsec\, respectively \citep[][]{Million16}. 
These parameters were chosen to effectively detect short duration UV flares \citep[e.g.][]{Brasseur19} and \textcolor{black}{were} noted by \citet{Million16} as providing a good midpoint in measurement error between the shortest and longest \galex\ integrations. We masked all UV fluxes with non-zero quality flags, to avoid systematic signals in our \galex\ lightcurves due to pixels contiguous to masked hotspots or sources observed near to the detector edge, which are prone to artifacts and diminished photometric performance \citep[][]{Million16}.

\section{Methods}
\textcolor{black}{In this section we describe the framework we have developed for testing the UV predictions of flare models calibrated using white-light observations. In Sect.\,\ref{sec:tess_flare_detect} and Sect.\,\ref{sec:galex_flare_detect_method} we describe our methods of detecting flares in \tess\ and \galex\ lightcurves and calculating flare energies. In Sect.\,\ref{sec:model_testing} we describe each of the flare models we have tested. In Sect.\,\ref{sec:tess_flare_occ_method} and
Sect.\,\ref{sec:galex_inject} we describe how we fit the average bolometric flare rates and predicted the UV flare activity for each sample. We describe how we compared the predicted UV flare rate for each model with the observed behaviour. In Sect.\,\ref{sec:method_correction} we detail how we used our results to calculate energy correction factors that can be used to bring the UV energy prediction of a chosen flare model in line with observations.} 

\subsection{\tess\ Flare Detection and Energy Calculation} \label{sec:tess_flare_detect}

To detect white-light flares in the \tess\ observations we first detrended the lightcurves following a method similar to that used in \citet{Jackman21kepler}. This method is based on those used for \kepler\ short cadence observations \citep[e.g.][]{Yang17} \textcolor{black}{and} uses a median filter to remove flares and isolate the quiescent stellar flux. 
To determine the size of the window for the median filter 
we first performed a generalised Lomb Scargle analysis of the lightcurve, searching for periods between 100 minutes and 10 days \textcolor{black}{and normalising with the residuals of the weighted mean of the \tess\ light curve} \citep[][]{Lomb76,Scargle82,Zechmeister09}. 
If the best fitting period had a power in the generalised Lomb Scargle periodogram greater than 0.25, it was selected. A power limit of 0.25 was selected \textcolor{black}{empirically} to avoid choosing window sizes based on false positive periods \citep[e.g.][]{Oelkers18}. 
We chose the window size for our median filter to be one-tenth the best fitting period, with limits of 30 minutes and 12 hours. These limits were chosen to avoid overly smoothing lightcurves and flares at short periods, and to avoid excessively large window sizes. If the power was less than 0.25, indicating either low-amplitude or non-detectable modulation, then a window size of six hours was automatically selected.

Each \textcolor{black}{\tess\ } lightcurve was 
split into continuous segments, separating on gaps 
of greater than 6 hours. Within each continuous segment we applied a median filter with our chosen window size. We used filters of diminishing window sizes at the edges of the lightcurve segment in order to preserve the signal in these regions. The lightcurve segment was then divided by the smoothed version. We calculated the mean and standard deviation, $\sigma$, of this resultant lightcurve. 3$\sigma$ outliers in the resultant lightcurve were then masked and interpolated over in the original lightcurve segment. This process was repeated until there were either no more recorded outliers, or it had run 20 times. We then divided the original lightcurve segment by the final smoothed version to create a detrended lightcurve segment.

To detect flares in an individual detrended lightcurve segment we calculated the median and the median absolute deviation (MAD) \textcolor{black}{the lightcurve segment}. We chose the MAD instead of the standard deviation for our flare detection as it is robust against outliers such as those from flares. 
To find flares we searched for consecutive outliers lying six MAD above the median of the lightcurve segment. Regions with at least two consecutive outliers six MAD above the median were flagged as flare candidates. Once the process was completed for all the segments within an individual lightcurve, we verified flare candidates through visual inspection \textcolor{black}{of both the lightcurves and \tess\ pixel files \citep[e.g.][]{Jackman21kepler,Vasilyev22}}. This was done to remove false positive signals, such as due to asteroid crossings in the \tess\ postage stamp, or candidate detections due to other astrophysical variablity (e.g. RR Lyrae). \textcolor{black}{These steps removed approximately 35 per cent of our flare candidates.} During this visual inspection stage we also manually set the start and end point of each flare. This was to correct for flares with very fast impulsive rises and decay phases which were also followed by longer gradual decays. In these cases, the automatic detection would flag the impulsive rise and initial decay only. By manually setting the end point we ensured that we calculated the full energy of every detected flare. 

To calculate the bolometric energies of white-light flares observed with \tess\ we followed the method outlined by \citet{Shibayama13}. This method assumes the flare spectrum can be modelled with a 9000\,K blackbody and equates the ratio of the star and flare luminosities within a given filter to the observed flare-only flux $\Delta F/F_{q}$, where $F_{q}$ is the quiescent flux. The renormalised blackbody is then integrated over all wavelengths to give the bolometric energy. We obtained the flare amplitude, $\Delta F/F_{q}$, for each flare by fitting a linear baseline to fluxes in the 20 minutes preceding and following each flare and subtracting it from the observed signal. This method assumes that any lightcurve modulation in the quiescent flux have timescales longer than the flare and can be fit with a line.

\subsection{\galex\ Flare Detection and Energy Calculation} \label{sec:galex_flare_detect_method}
To search for flares in the \galex\ lightcurves we followed a method adapted from the one used by \citet{Brasseur19}. 
We outline this method here for a single lightcurve. We first searched for data points lying at least 3.5$\sigma$ above the global median of the lightcurve, where $\sigma$ \textcolor{black}{here} is the uncertainty of each observation. We checked each flagged point to make sure there was at least one adjacent point lying 2$\sigma$ above the global median. If there was no such adjacent point, the flagged 3.5$\sigma$ outlier was removed as a flare candidate in our search. The difference between the maximum flux of each flare candidate and the global median was also required to be greater than the difference between the global minimum and the global median. 

We then split the lightcurve into continuous regions. Each continuous region was separated by at least 1600s from another. For each flare candidate, we isolated the continuous region it was in to calculate the flare edges. For a given candidate, the \textcolor{black}{maximum of the } 3.5$\sigma$ outliers was considered to be the flare peak. The start and end of the flare were where the lightcurve first went below the global median, or the edge of the continuous region, whichever came first. 

We then visually inspected each flare candidate in order to remove false positives. Signals that can cause false positive detections include large scale periodicity and instrumental effects \citep[e.g.][]{Million16}. \citet{Brasseur19} found that these signals could sometimes dominate individual visits and obscure potential flare events. To filter their sample, they automatically excluded any flare candidate that lasted an entire visit. However, while this removed such false positive signals, it also removed true high energy flare events that dominate a given visit. These events are important for this study, in particular when we run flare injection tests in Sect.\,\ref{sec:galex_inject}. We therefore kept these signals prior to our visual inspection. \textcolor{black}{Our visual inspection removed 43 and 24 per cent of the \galex\ NUV and FUV flare candidates respectively. \citet{Brasseur19} removed 53 per cent of their NUV flare candidates after visual inspection, however also removed events that dominated a single visit.}

We calculated flare energies in the \galex\ NUV and FUV bandpasses following the method of \citet{Brasseur19}. We first subtracted the quiescent flux from a given flare. The quiescent flux was calculated either using the median of the flux preceding the flare, or the median of the entire lightcurve if a flare dominated a visit.
The energy in a given filter, $E_{NUV,FUV}$, is then calculated using
\begin{equation}
    E_{NUV,FUV} = 4\pi d^{2}\Delta \lambda_{NUV,FUV}\int_{t_{start}}^{t_{stop}}\,F(t) dt
\end{equation}
where $d$ is the distance to the star in cm, $\Delta\lambda_{NUV,FUV}$ is the equivalent width (FWHM) of the chosen filter in \angstrom\ and $F(t)$ is the quiescent subtracted flare flux from the start of the flare $t_{start}$ to the end $t_{end}$. The FWHM of the NUV filter is \textcolor{black}{795.65}\,\angstrom\ and the FUV filter is \textcolor{black}{227.81}\,\angstrom\ \citep[][]{Rodrigo12,Rodrigo20}. As \galex\ provided flux-calibrated time-tagged data, this method calculates the flare energy in a way that is independent of a chosen model.

\subsection{Testing empirical flare models with \tess\ and \galex\ } \label{sec:model_testing}
\begin{figure}
    \centering
    \includegraphics[width=\columnwidth]{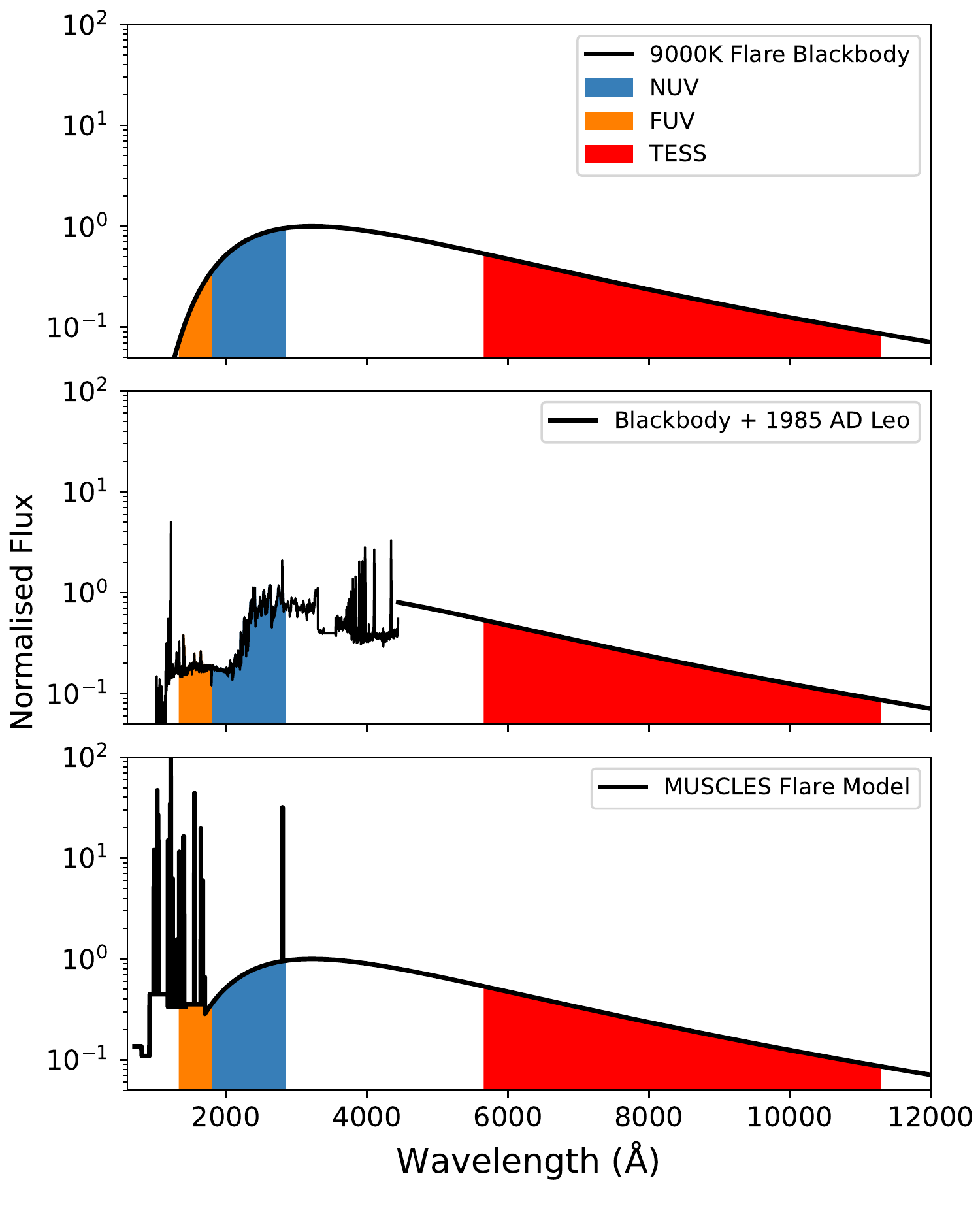}
    \caption{The three types of flare models tested in this work. The orange, blue and red sections indicate the wavelength coverage of the \galex\ FUV, NUV and the \tess\ bandpasses respectively. Each model has been normalised to give the same amount of flux in the \tess\ bandpass. The top panel shows the 9000\,K blackbody model typically used in flare studies to calculate bolometric energies. The adjusted blackbody model (Sect.\,\ref{sec:model_adjusted}) applies a multiplicative factor to the UV fluxes predicted by this model to account for emission lines. The middle panel shows a representative view of the 9000\,K blackbody plus the 1985 AD Leo flare spectrum \citep{Hawley91,Segura10}, assuming the original spectrum but a U band energy of $10^{34}$ erg (see Sect.\,\ref{sec:model_ad_leo}). 
    Here we used a $U$ band energy fraction of 0.11 from \citet{Glazier20}, corresponding to the third AD Leo Great Flare model in Tab.\,\ref{tab:uv_fractions}. The bottom panel shows the MUSCLES flare model from \citet{Loyd18muscles}, which combines a 9000\,K blackbody continuum with NUV emission lines and an FUV flare spectrum based on \hst\ observations.}
    \label{fig:flare_models}
\end{figure}

The presence of data from both \tess\ and \galex\ for a set of stars enables us to test the UV predictions of various empirical flare models. These models cover optical, NUV and FUV wavelengths. Once \textcolor{black}{it is} calibrated with optical data, \textcolor{black}{a flare model} can offer a way of predicting the UV flaring activity of low-mass stars. We tested six models in this work and we discuss these below. We calculated the fraction each model emits in the \galex\ NUV and FUV bandpasses, and the ratio of each to that for a 9000\,K blackbody model. These values are shown in Tab.\,\ref{tab:uv_fractions}. We plotted the three main types of models in Fig.\,\ref{fig:flare_models}.

\subsubsection{9000K Blackbody} \label{sec:model_9000}
The first model we tested is the 9000\,K blackbody spectrum. This model is based on optical multi-colour photometry and spectroscopy of flares from active M dwarfs \citep[e.g.][]{Hawley92, Kowalski13}. 
This model represents the most basic description of the white-light and UV emission and forms the basis of other flare models. 
It lacks any contribution from emission lines such as the Balmer series in the optical and species such as Mg, Fe and Si in the UV. 
\citet{Kowalski13} measured from optical flare spectra that the Balmer jump increased the continuum flux by between factors of 1.3 and 4.3, with the exact contribution appearing to depend on the impulsivity and temperature of the flare. 
\citet{Kowalski19} found this model underestimates the NUV flux at the flare peak by a factor of 3 when compared to \hst\ NUV observations of two flares from GJ 1243, due in part to the elevated Balmer continuum stretching into NUV wavelengths.  \citet{Kowalski10} and \citet{Davenport12} combined a $\sim10,000$\,K thermal blackbody with the radiative-hydrodynamic (RHD) simulations from \citet[][]{Allred06} to create two-component flare models that combined a $10,000$\,K thermal blackbody with varying strengths of Balmer jump. They used this for modelling the optical continuum, but stopped short of the \galex\ NUV bandpass. We discuss the use of the 9000\,K blackbody with a second component in Sect.\,\ref{sec:discussion_flare_teff}.

To calculate the fraction of energy this model emits within the \galex\ NUV and FUV passbands, we integrated the blackbody within the wavelengths covered by the \galex\ NUV and FUV filters and divided these values by bolometric flux of the blackbody \citep[][]{Brasseur19}. 
We calculated that the 9000K blackbody model emits 15.2 and 1.8 per cent of its flux in the \galex\ NUV and FUV bandpasses, respectively.

\subsubsection{Adjusted 9000K Blackbody} \label{sec:model_adjusted}
The second model is the ``adjusted blackbody'' model. This model is the 9000\,K blackbody spectrum multiplied by a \textcolor{black}{scaling} factor to incorporate the flux from emission lines. Previous studies \citep[e.g.][]{Glazier20} have used the ratio of the bolometric flare energy to the continuum-only energy, 0.6, from \citet{Osten15}. This value was calculated by combining archival multi-wavelength observations of flares from the active M dwarf AD Leo from \citet{Hawley91} and \citet{Hawley95}. \citet{Osten15} specified the bolometric energy in their study as the sum of the optical and UV emission from the photosphere and chromosphere, and the high energy EUV and X-ray emission associated with the corona, as opposed to the total energy from the flare footpoints only. 
\citet{Osten15} calculated, using values from \citet{Hawley91}, that a 9000-10,000\,K continuum emits 90 per cent of the \textcolor{black}{flare radiated energy in the optical and UV. The adjusted blackbody model} predicts 1.11 times the UV fluxes of the continuum-only 9000\,K blackbody model \textcolor{black}{to include the optical and UV flux contribution from emission lines.}

\subsubsection{9000\,K Blackbody plus the 1985 AD Leo Flare} \label{sec:model_ad_leo}
The third set of models we tested are based on spectra of the 1985 ``Great Flare'' from the active M dwarf AD Leo \citep[][]{Hawley91}.
\citet{Hawley91} observed this flare spectroscopically in the UV with the \textit{International Ultraviolet Explorer} (\iue) and in the optical with ground-based observations. 
\citet{Hawley91} calculated that this flare had a U band energy of $10^{33.8}$ ergs, which subsequent works have approximated to $10^{34}$ ergs \citep[][]{Rimmer18}.  \citet{Segura10} used the time-resolved spectra to construct a time-dependent flare model, which they used to study \textcolor{black}{the effect of high energy flare on the atmosphere of an Earth-like exoplanet.} \citet{Rimmer18} later used this model to study the effects of NUV flare flux on the habitability of rocky exoplanets around low-mass stars. \textcolor{black}{They used their results to define a region in flare frequency space called the ``abiogenesis zone''. } This is the region on flare frequency diagrams where flares occur frequently enough to provide the NUV flux required for prebiotic chemistry to be viable on \textcolor{black}{the surfaces of} rocky planets around M stars.

\textcolor{black}{During their analysis }\citet{Rimmer18} noted that the relatively flat spectrum of the 1985 AD Leo flare meant that the number of NUV photons deposited was linearly proportional to the energy in the U band, and the NUV photon flux and abiogenesis zone could be \textcolor{black}{predicted} from the U band energy. Assuming the 1985 flare had a U band energy of $10^{34}$ erg, \citet{Rimmer18} calculated a relation between the U band flare energy and NUV photon flux of the AD Leo flare. 
Studies of flares from low-mass stars have since used the U band energy to NUV flux relation from \citet{Rimmer18} to test whether specific stars flare enough to make prebiotic chemistry viable. 
One method of doing this is to calculate bolometric flare rates from white-light flare observations, assuming a 9000\,K blackbody, and multiplying these rates by the fraction the 9000\,K blackbody emits within the U band to calculate a U band flare rate  \cite[e.g.][]{Gunther20,Ducrot20,Glazier20}. These studies found that only the most active stars in their sample flared with rates that could reach the abiogenesis zone. \citet{Ducrot20} and \citet{Glazier20} used separate observations in the near-IR (with TRAPPIST) and the optical (with EvryScope and \ktwo) to determine that TRAPPIST-1 does not flare enough for prebiotic photochemistry to be viable on the surfaces of the orbiting exoplanets.

To test this model 
we recreated the time-dependent 1985 AD Leo flare model from \citet{Segura10}. We integrated this model in the wavelengths covered by the \galex\ NUV and FUV bandpasses to calculate the corresponding \galex\ UV energies. These were $7.3\times10^{33}$ and $1.1\times10^{33}$ erg respectively. We did not use the UV energies reported by \citet{Hawley91} due to the differences between the wavelengths covered by the \iue\ and \galex\ UV bandpasses. However, we note that the \galex\ bandpasses are both covered entirely by the \iue\ spectra. Our calculated \galex\ NUV and FUV energies are 73 and 11 per cent of the $10^{34}$ erg U band energy assumed by \citet{Segura10} and \citet{Rimmer18}. To calculate the fraction of UV energy relative to the bolometric energy of a 9000\,K blackbody, we multiply these values by the fraction emitted by this blackbody in the U band. 

Different studies have calculated different values for the fraction of flare energy emitted in the U band. \textcolor{black}{\citet{Gunther20} and \citet{Ducrot20} multiplied the 9000\,K blackbody with the spectral response of the U band and integrated over the result to calculate U band fractions of 7.6 and 6.7 per cent respectively. }
\citet{Glazier20} used the U band to bolometric energy ratio of 11 per cent from \citet{Osten15} for their analysis of the TRAPPIST-1 system. To test these values, we generated three separate versions of this model (referred to as submodels), with each one using the respective U band fraction to estimate the UV fluxes. 
The first submodel uses the \citet{Ducrot20} U band energy fraction of 6.7 per cent, which corresponds to \galex\ NUV and FUV fractions of 4.9 and 0.7 per cent the energy of the 9000\,K blackbody respectively. The second submodel uses the \citet{Gunther20} value of 7.6 per cent, which corresponds to \galex\ NUV and FUV fractions of 5.5 and 0.84 per cent respectively. The third submodel uses the \citet{Glazier20} value of 11 per cent, which corresponds to \galex\ NUV and FUV fractions of 8.0 and 1.2 per cent respectively. 
The ratios between these fractions and the fraction emitted by the 9000\,K blackbody are shown in Tab.\,\ref{tab:uv_fractions}.

\begin{table*} 
    \centering
    \begin{tabular}{c|c|c|c|c}
    \hline
         Model Number & Model Name &  $\mathrm{NUV_{model}/NUV_{9000}}$  & $\mathrm{FUV_{model}/FUV_{9000}}$ \\
         \hline
         1 & 9000\,K blackbody & 1.00 & 1.00 \\
         2 & Adjusted blackbody & 1.11 & 1.11 \\
         3 & AD Leo Great Flare, 1 & 0.32 & 0.39 \\ 
         4 & AD Leo Great Flare, 2 & 0.36 & 0.47 \\
         5 & AD Leo Great Flare, 3 & 0.53 & 0.66 \\
         6 & MUSCLES Model & 1.13 & 3.33  \\ 
         \hline
    \end{tabular}
    \caption{The six models used in this work to predict the \textcolor{black}{\galex} NUV and FUV fluxes. We have provided the ratio of the \textcolor{black}{\galex} UV emission of each model and the UV emission of the 9000\,K blackbody. Each model has been normalised to give the same amount of flux as the 9000\,K blackbody in the \tess\ bandpass. The 9000\,K blackbody emits 0.152 and 0.018 of its flux in the \galex\ NUV and FUV bands respectively.}
    \label{tab:uv_fractions}
\end{table*}

\subsubsection{MUSCLES Flare Model} \label{sec:model_muscles}
The fourth model is the optical+UV MUSCLES flare spectrum from \citet{Loyd18muscles}. The original version of this model uses a 9000\,K blackbody for the optical emission and the NUV continuum. The NUV emission includes additional flux from empirically defined NUV Mg II h\&k emission lines. The FUV emission is an empirical model developed from the energy budget of individual emission lines. These energies were measured from \hst\ spectra of flares from the MUSCLES survey of M stars. 
By using the measured flare energy within the \tess\ bandpass to calibrate the continuum emission from the 9000\,K blackbody, this model can be used to predict UV flaring activity directly from \tess\ observations while incorporating the energy contribution from emission lines. This model was used by \citet{Chen21} to calculate the FUV energies of flares detected with \tess\ and to study their effects on exoplanet atmospheres.

This model has the highest NUV and FUV emission of all the tested models. This is due to the contribution of the Mg II h\&k lines in the NUV, and both the line and elevated continuum emission in the FUV. The model presented in \citet{Loyd18muscles} uses a temperature of 9000\,K for the blackbody, however this can be changed at the user's preference. We chose to keep the flare temperature as a 9000\,K blackbody during our testing, for consistency with the other models tested in this work. To calculate the UV energy relative to the bolometric energy from a lone 9000\,K blackbody, we normalised this model so that the flux from the continuum emission in the \tess\ bandpass matched that of the 9000\,K blackbody. We then calculated the integrated flux in the wavelengths covered by the \galex\ UV bandpasses, and divided this by the bolometric flux of the lone 9000\,K blackbody. This gave \textcolor{black}{\galex} NUV and FUV emission fractions of 17.4 and 6 per cent, respectively. This was 1.13 and 3.33 times the \textcolor{black}{\galex} NUV and FUV fluxes of the 9000\,K blackbody model alone.  

\begin{figure}
    \centering
    \includegraphics[width=\columnwidth]{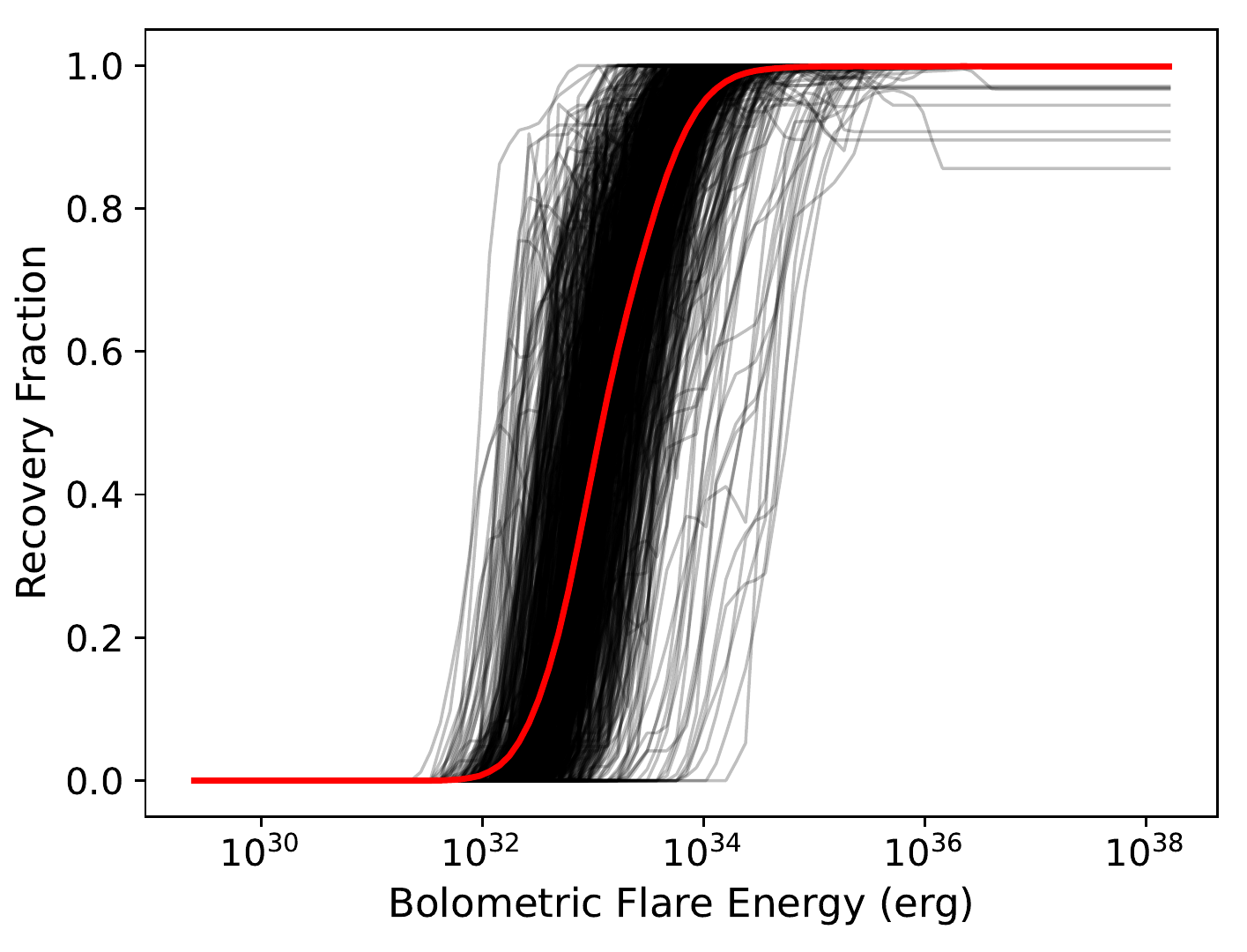}
    \caption{Our measured recovery rate for stars in our partially convective M star sample. Each grey line is an individually measured recovery rate, while the red line is the measured rate $R(E)$ for the full sample. We calculated this following the method outlined in Sect.\,\ref{sec:tess_flare_occ_method}. At low energies we are unable to detect any flares, while at high energies we can detect every event. Note that the x axis is on a logarithmic scale, and how the detection efficiency does not scale linearly with energy.}
    \label{fig:recovery_rate}
\end{figure}

\begin{figure}
    \centering
    \includegraphics[width=\columnwidth]{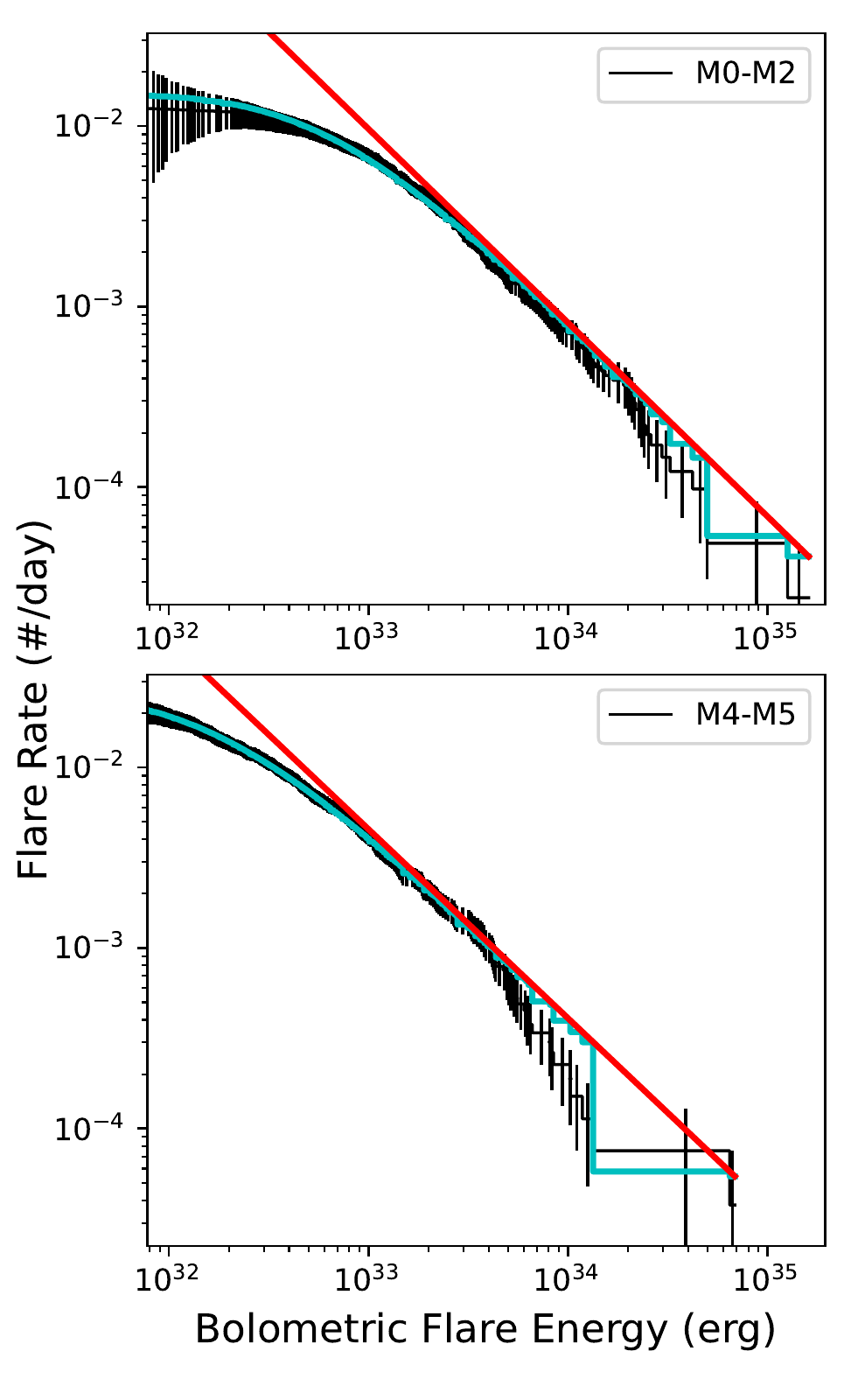}
    \caption{The average bolometric flare occurrence rates for the partially and fully convective M stars in our sample with \galex\ NUV observations. The black line is the measured average \tess\ flare rate calculated by combining both flaring and non-flaring stellar lightcurves. The cyan line is the fitted observed flare rate. The red line is the intrinsic flare power law, as measured before being convolved with the \tess\ detection method. The fitted power law values for these samples, and the corresponding FUV samples, can be found in Tab.\,\ref{tab:fitted_ffd}. }
    \label{fig:tess_flare_rates}
\end{figure}

\subsection{Measuring the \tess\ white-light flare occurrence rate} \label{sec:tess_flare_occ_method}
We used our \tess\ observations to measure the average white-light flare occurrence rate in each of our \textcolor{black}{mass and UV} subsets. 
We measured the average flare rates in each optical and UV sample to account for any changes in the underlying magnetic activity that may have occurred in the 6 to \textcolor{black}{17 year interval} between the \galex\ and \tess\ observations. 
Works using ground-based photometry and observations of emission lines such as H$\alpha$ and Ca II H\&K have found evidence for activity cycles in low-mass stars with periods of several years up to a decade or more in length \citep[e.g.][]{Buccino11,Robertson13,DiezAlonso19}. 
While such changes may affect the observed energies and flare rates between our two sets of observations for a single source, by measuring the average flaring behaviour of hundreds of stars they can be smoothed over. 

It has previously been shown that stellar flares occur with a power law distribution in energy, $E$ \citep[e.g.][]{Lacy76}. This distribution can be written as 
\begin{equation} \label{eq:edit_powlaw}
    \frac{dN(E)}{dE} = k E^{-\alpha} ,
\end{equation}
where $N(E)$ is the number of flares of a given energy $E$ that occur in a time period, $k$ is a constant of proportionality and $\alpha$ is the power law index \citep[][]{Audard00}. Flare studies generally do not measure the distribution in Eq.\,\ref{eq:edit_powlaw}, instead measuring the rate of observed flares with energy $E$ or greater. This distribution can be obtained by integrating Eq.\,\ref{eq:edit_powlaw} from energy $E$ to infinity, resulting in 
\begin{equation} \label{eq:ffd}
    \log_{10}{\nu} = C + \beta \log_{10}{E}
\end{equation}
where $\nu$ is the number of flares in a time period with an energy of $E$ or greater, $C = \log\big(\frac{k}{1-\alpha}\big)$ and $\beta= 1-\alpha$ \citep[e.g.][]{Hawley14}. By fitting Eq.\,\ref{eq:ffd} to the observed \textcolor{black}{flare rate}, studies can measure $C$ and $\alpha$ and calculate the predicted flare rate at a given energy.

However, observations often show a deviation from Eq.\,\ref{eq:ffd} at low energies \citep[e.g.][]{Pettersen84,Ramsay13,Gilbert21}, where the observed cumulative flare frequency turns over and flattens. This turnover is attributed to the decreasing efficiency of the flare detection method at low energies. 
One way studies have accounted for this turnover is to mask flares with energies below the turnover before fitting the observed flare rate \citep[e.g.][]{Hawley14}. This limit has previously been chosen as where the distribution is no longer consistent with a power law \citep[][]{Lin19}, or where the detection efficiency drops below some threshold as determined through flare injection and recovery tests \citep[e.g. 68 per cent;][]{Davenport16,JackmanOrion20}. 
Depending on the chosen threshold, these methods \textcolor{black}{may limit }their sample to the intrinsically rarer higher energy events. 
Along with this, near the chosen limit the detection method may still miss a non-trivial number of flares with energies close to the detection threshold, something that will result in a shallower measured occurrence rate and fitted power law distribution. 

\citet{Jackman21} presented a generalised method to fit the flare occurrence rate while accounting for the detection efficiency. In this analysis, the intrinsic flare distribution in Eq.\,\ref{eq:edit_powlaw} is multiplied by the detection efficiency in terms of energy, $R(E)$, and integrated in energy to give the observed FFD,  
\begin{equation} \label{eq:obs_ffd}
    N(E>E_{f}) = \frac{k}{\alpha-1} \Bigg(R(E_{f})E_{f}^{-\alpha + 1} +
    \int_{E_{f}}^{E_{max}}R'(E)E^{-\alpha+1}dE\Bigg)
\end{equation}
where $E_{f}$ is the observed flare energy and $E_{max}$ is the energy at which the chosen detection method saturates to its maximum efficiency. By fitting Eq.\,\ref{eq:obs_ffd} instead of Eq.\,\ref{eq:ffd}, we can use all the observed flares in our fitting, in particular those at low energies that would otherwise fall below a user-defined detection threshold.

We performed flare injection and recovery tests to measure the efficiency $R(E)$ of our flare detection method for each of our low-mass star subsets. We did this for all of the stars in each subset, both flaring and non-flaring, to in order to determine the average detection efficiency in each sample. We \textcolor{black}{followed} the method of \citet{Jackman21}, who measured the average flare occurrence rates of K and M dwarfs observed with NGTS. This method is \textcolor{black}{itself} based on injection and recovery techniques presented in \citet{JackmanOrion20} and \citet{Davenport16}. 
\textcolor{black}{We first} measured the detection efficiency of each star. We generated flares using the \citet{Davenport2014} empirical flare model for each lightcurve. This flare model was generated from 1 minute cadence \kepler\ observations of white-light flares from GJ 1243. We split each \tess\ lightcurve into continuous segments. We injected 100 simulated flares into each continuous segment and ran the detrending and flare detection methods outlined in Sect.\,\ref{sec:tess_flare_detect}. The properties of the generated flares were drawn randomly from uniform distributions between 0.01 and 10 times the quiescent flux for the flare amplitude and 
2 and 70 minutes for the flare full-width at half-maximum (FWHM). These values were chosen to be representative of the amplitudes and durations of flares 
detected in \tess\ lightcurves \citep[e.g.][]{Gunther20}. We then tested whether each flare was successfully detected, assigning a value of 1 if it was, or 0 if it was not. We then calculated the energy of each injected flare following the method outlined in Sect.\,\ref{sec:tess_flare_detect}. We then used these energies and the results of our detection tests to measure the efficiency of our detection method. We used 20 bins spaced logarithmically in energy and calculated the ratio of successfully recovered flares in each bin. Finally for a single star, we used a Wiener filter of three bins to smooth the recovery fraction, following \citet{Davenport2014}. This was performed for every star in our sample. 

\textcolor{black}{In order to measure the average flare detection efficiency for use in Eq.\,\ref{eq:obs_ffd} we must account for the different duration each star might have been observed for, and the varying brightness of each star. }
In order to do this we considered our measurement of the average flare rate from a sample of stars to be equivalent to observing one star that is representative of the average behaviour of the sample. \textcolor{black}{This average star} was observed for a time equal to the combined observing duration of the chosen sample. \textcolor{black}{To calculate the average detection efficiency we first }interpolated the recovery fraction for each individual star onto a single grid in energy. This energy grid was spaced logarithmically in 100 steps from one order of magnitude below our lowest measured flare energy to one magnitude above our highest measured flare energy. Once each recovery fraction was on the same energy grid, we multiplied them by their respective observing duration, to determine an ``equivalent observing time'' that each star observed flares of a given energy for. These new recovery fractions were then summed and divided by the total observing duration for the chosen subset. This resulted in a single recovery fraction that ranged between 0 and 1. It was this recovery fraction that we used in Eq.\,\ref{eq:obs_ffd} to fit the observed average flare occurrence rate for each subset. An example of the measured recovery fraction for the partially convective M star sample is shown in Fig.\,\ref{fig:recovery_rate}.

To fit the average flare occurrence rate for each subset, we used a Markov Chain Monte Carlo (MCMC) process. We generated the MCMC process using the {\scshape emcee} Python package \citep[][]{Foreman-Mackey13}. We used 32 walkers for 10,000 steps, using the final 2000 steps to sample the posterior distribution. We assumed Poisson errors for the observed flare occurrence rate. However, prior to fitting we multiplied these errors in the observed \textcolor{black}{flare rate} by $R(E)^{-1/2}$ to account for possible uncertainties in our flare recovery testing for the smallest flares \citep[e.g.][]{Ilin19}. By fitting the observed \textcolor{black}{flare rate} while simultaneously accounting for the detection efficiency, we use our fitted values of $k$ and $\alpha$ to retrieve the intrinsic \textcolor{black}{flare rate}. We performed our fitting, along with the associated injection and recovery tests, for the flare rates calculated using the 9000\,K blackbody model. The results of our fitting for the NUV subset of partially and fully convective M stars are shown in Fig.\,\ref{fig:tess_flare_rates}, and the fitted power law values for all subsets are given in 
Tab.\,\ref{tab:fitted_ffd}. We note that the average flare rate for our fully convective M star sample in Fig.\,\ref{fig:tess_flare_rates} appears to drop below that expected for a power law distribution at around $10^{34}$erg, although is within $2\sigma$ of our observations and returns within $1\sigma$ at higher energies. To confirm that we were not underestimating the energies of these flares, we manually rechecked the start and end times of each event. We found that all energies were calculated for the full observed flare duration and noted no issues with our baseline subtraction when calculating the flare energy. Deviations such as this have been observed in previous studies that have measured average flare rates \citep[e.g.][]{Howard19,Ilin21}, suggesting that this drop may be a real feature. One possibility is that some of these flares are superpositions of multiple events \citep[e.g. from sympathetic flaring;][]{Moon02}. Such superpositions would appear rarer than single events, but their energy may not be high enough to match the predicted value at the observed rate. We note that our fit matches the energies above and below this region, so we have used our results in the rest of this work. 

\begin{table*}
    \centering
    \begin{tabular}{c|c|c|c|c|c|c|c|c|}
    \hline
    Subset & $\log k$ & $\alpha$ & $C$ & $N_{flares}$ & $N_{stars, flare}$ & $N_{stars, total}$ & \textcolor{black}{$N_{flares, \galex\ NUV}$}  \tabularnewline \hline
    \begin{tabular}{@{}c@{}} \tess\ observations of stars\\ with \textcolor{black}{\galex} NUV coverage\end{tabular}
    & & & & &  &  & &  
    \tabularnewline \hline
    M0V-M2V & $33.5\pm0.2$ & $2.07\pm0.01$ & $33.4\pm0.2$ & 514 & 69 & 758 & 7 \tabularnewline
    M4V-M5V & $32.2\pm0.2$ & $2.05\pm0.01$ & $32.2\pm0.2$ & 676 & 107 & 492 & 14 \tabularnewline
    \hline
    \begin{tabular}{@{}c@{}} \tess\ observations of stars\\ with \textcolor{black}{\galex} FUV coverage\end{tabular}
    &  &  & &  & & & \textcolor{black}{$N_{flares,\galex\ FUV}$}
    \tabularnewline \hline
    M0V-M2V & $33.4\pm0.3$ & $2.07\pm0.01$ & $33.4\pm0.3$  & 389 & 49 & 484 & 1 \tabularnewline
    M4V-M5V & $32.6\pm0.4$ & $2.06\pm0.01$ & $32.6\pm0.4$ & 347 & 73 & 312 & 4 \tabularnewline 
    \hline
    \end{tabular}
    \caption{The results of our fitting to the average \textcolor{black}{bolometric} flare rates \textcolor{black}{measured from \tess\ 2-minute cadence photometry}. We measured separate bolometric flare rates for the \textcolor{black}{samples of stars with NUV and FUV coverage,} 
    due to the decreased number of stars with FUV data. We fit each sample following the method outlined in Sect.\,\ref{sec:tess_flare_occ_method}. \textcolor{black}{We have also provided the number of flares detected with \galex\ for each sample.}}
    \label{tab:fitted_ffd}
\end{table*}

\subsection{Using The \tess\ Flare Rate To Predict The UV Activity} \label{sec:galex_inject}
\textcolor{black}{We calculated the bolometric energies in Sect.\,\ref{sec:tess_flare_detect} and flare rates in Sect.\,\ref{sec:tess_flare_occ_method} for each mass and \textcolor{black}{\galex} UV sample assuming} the flare spectrum was represented by a 9000\,K blackbody. 
To calculate the predicted \textcolor{black}{\galex} UV activity \textcolor{black}{for this model }we renormalised the \textcolor{black}{fitted bolometric} flare rate, assuming the slope of the power law $\alpha$ does not change \textcolor{black}{significantly} between the bolometric and \textcolor{black}{\galex} UV rates \citep[e.g.][]{Maehara15}. Therefore, only the normalisation constant $C$ will change. The value of \textcolor{black}{$C_{UV}$} was calculated using 
\begin{equation} \label{eq:renormalisation_model}
    C_{UV} = C_{Bol} - \beta\log_{10}{f}
\end{equation}
where $C_{Bol}$ is the normalisation constant for the \textcolor{black}{bolometric} flare rate and $f$ is the ratio between the \textcolor{black}{bolometric and \textcolor{black}{\galex} UV} energies \textcolor{black}{for the 9000\,K blackbody model. These were 0.152 and 0.018 for the \galex\ NUV and FUV bandpasses respectively. To calculate the UV flare rate for the other models, we adjusted the coefficients in Eq.\,\ref{eq:renormalisation_model} for use with the \textcolor{black}{\galex} UV flare rate from the 9000\,K blackbody model. The adjusted relation was}
\begin{equation} \label{eq:renormalisation_model_uv}
    C_{UV, model} = C_{UV, 9000K} - \beta\log_{10}{f_{model}}
\end{equation}
\textcolor{black}{where $C_{UV, model}$ was the normalisation constant of the chosen model, $C_{UV, 9000K}$ was the normalisation constant of the 9000\,K blackbody ($C_{UV}$ in Eq.\,\ref{eq:renormalisation_model}) and $f_{model}$ is the ratio between the \textcolor{black}{\galex} UV emission of a chosen model and the 9000\,K blackbody from Tab.\,\ref{tab:uv_fractions}.}  

Due to the irregular time sampling and short visit windows of the \galex\ observations for individual stars, we are not able to test each model by directly comparing their predicted \textcolor{black}{\galex} UV flare rate with the measured values. 
As \galex\ only observed for up to \textcolor{black}{30 minutes} per visit, some flares flagged in our analysis in Sect.\,\ref{sec:galex_flare_detect_method} may have only been partially observed, resulting in lower measured energies. This effect can also complicate the measurement of the recovery fraction $R(E)$ for the \galex\ data, due to the degeneracy between injected and the measured energies of partially observed flares. Therefore, to compare our predicted and observed \textcolor{black}{\galex} UV flare rates while accounting for the effects of the \galex\ observation strategy, we performed flare injection and recovery tests for each model to calculate artificial observed flare rates. By performing such tests we could incorporate the sampling of the \galex\ observations and in turn measure the predicted observed average \textcolor{black}{\galex} UV flare rates, which could be compared with the actual observed flare rates we measured.

We used the calculated \textcolor{black}{\galex} NUV or FUV flare occurrence rate to generate simulated ``flare-only'' lightcurves for each model and star in our sample. These flares had energies drawn from the predicted UV flare rates. We designed each lightcurve to extend from three hours before to three hours after each \galex\ visit. The peak times of injected flares were placed randomly throughout this time span. 
We initially generated a grid of 10,000 flares using the \citet{Davenport2014} \textcolor{black}{empirical flare} model. These flares had amplitudes chosen randomly from a uniform distribution between 0.1 and 100 and FWHM chosen randomly from a uniform distribution between 10 seconds and \textcolor{black}{5} minutes. The amplitude and FWHM values were \textcolor{black}{based on previous NUV and FUV observations of flares with \galex\ \citep[e.g.][]{Welsh07,Brasseur19} and \hst\ \citep[e.g.][]{Loyd18,Loyd18muscles,Kowalski19}.} 
\textcolor{black}{\citet{Loyd18,Loyd18muscles} measured FWHM values of FUV flares from M stars of between 10 and approximately 140 seconds.} 
\citet{Brasseur19} measured full flare durations up to 5 minutes for low energy NUV flares observed with \galex\ \textcolor{black}{and noted that this upper limit did not appear to strongly depend on the duration of an individual \galex\ visit.} 
\textcolor{black}{We chose an upper FWHM of 5 minutes to make sure our injected flares fully encompassed the full observed range of \textcolor{black}{\galex} UV flare FWHM durations. We also did this to include longer duration flares that may have been missed in previous studies due to the short visit and orbit durations offered by \galex\ and \textit{HST} in comparison to optical studies.}

We calculated the energy of each artificial flare in our grid using the method outlined in Sect.\,\ref{sec:galex_flare_detect_method}. We then picked flares from the grid with energies that would satisfy our calculated UV occurrence rate for the \galex\ observations. \textcolor{black}{Flares were placed randomly throughout a simulated lightcurve. This simulated lightcurve was then interpolated onto the times of the \galex\ observations. As we simulated times before, during and after each \galex\ visit, this interpolation will recreate flares which were partially observed. The simulated flare-only lightcurve was added to the quiescent-only \galex\ UV lightcurve to create a UV lightcurve with injected flares.} To avoid our detection algorithm triggering on events other than the injected flares, we masked all events that were flagged in our initial flare search. 

\textcolor{black}{We performed our injection and recovery tests 10,000 times per star for each model. We did for all six models, for both partially and fully convective stars, in the \textcolor{black}{\galex} NUV and FUV. We chose 10,000 times to }
fully explore how the \galex\ observing strategy impacts the observed flare rate. We then used the results of these tests from all stars in the chosen sample to measure 10,000 average \textcolor{black}{\galex} UV flare rates. Each flare rate was interpolated onto the same grid in energy. We then calculated the 16th, 50th and 84th percentile at each step in energy, and used these to calculate the median predicted observed flare rate and the lower and upper \textcolor{black}{uncertainties} at each step in energy.

To \textcolor{black}{compare the average flare rate each model predicted we should observe, and the rate we actually did observe, }
we assumed that every UV flare has an optical counterpart \citep[e.g.][]{Fletcher07, Kowalski19}. This assumption allowed us to calculate the discrepancy in energy, at a given flare rate, for each model. We discuss the validity of this assumption in Sect.\,\ref{sec:inferring_activity}. \textcolor{black}{These values are measures of the apparent difference in energy after selection effects due to the \galex\ observing strategy and should not be used to correct \textcolor{black}{\galex} UV predictions of flare models (see Sect.\,\ref{sec:method_correction}).}
For each scenario, \textcolor{black}{we calculated the difference in energy between the predicted and observed flare rates. When we could not retrieve enough flares from our injection and recovery tests to measure a predicted flare rate, we calculated a lower limit.}

\subsection{Calculating UV Energy Correction Factors} \label{sec:method_correction}
An important part of this work is to calculate energy correction factors (ECFs) for each model. These ECFs can be used to bring the predicted \textcolor{black}{\galex} UV flare energy for each model in line with observations. 
To calculate these correction factors we performed a second round of flare injection and recovery tests for each stellar subset. For each mass subset we injected flares with \textcolor{black}{\galex} UV power laws equivalent to having energies up to \textcolor{black}{50} times that of the 9000\,K blackbody. 
We then ran injection and recovery tests for each new flare rate as discussed in Sect.\,\ref{sec:galex_inject}, performing our tests 10,000 times per model and mass subset. We did this for both the \galex\ NUV and FUV bandpasses.

\textcolor{black}{We used the results of our new set of injection and recovery tests to generate} 
a grid of predicted average \textcolor{black}{\galex} UV flare rates. By interpolating over this grid we were able to determine the predicted observed \textcolor{black}{\galex} UV flare rate for any given predicted UV energy fraction. We ran an MCMC process to measure the best fitting \textcolor{black}{\galex} UV energy fraction, and thus the correction factor for each model. As for measuring the average white-light flare rate in Sect.\,\ref{sec:tess_flare_occ_method}, we used 32 walkers for 10,000 steps, using the final 2000 steps to sample the posterior distribution. To account for the uncertainty in our predicted observed \textcolor{black}{\galex} UV flare rates, we used the results of each of our 10,000 flare injection and recovery tests directly in our fitting. At each step of the MCMC process we randomly selected a test and used its result to re-generate the grid of predicted \textcolor{black}{\galex} UV flare rates for each UV energy fraction. 

We did not use the results of the initial flare injection and recovery tests for each model \textcolor{black}{from Sect.\,\ref{sec:galex_inject} }to calculate correction factors. This is because the efficiency of our flare detection method can change non-linearly with energy. An example of this is shown in  Fig.\,\ref{fig:recovery_rate} for our \tess\ white-light data. 
If this is not accounted for, calculated correction factors may result in corrected models overestimating the predicted \textcolor{black}{\galex} UV flare energies. By running a series of tests with \textcolor{black}{\galex} UV rates that vary prior to flare injection, we can account for the changing \galex\ flare detection efficiency and calculate more accurate correction factors. 
To calculate the correction factors \textcolor{black}{for our other models, we divided the measured factor for the 9000\,K blackbody model by the \textcolor{black}{\galex} UV ratios given in Tab.\,\ref{tab:uv_fractions}.}

\section{Results} \label{sec:results}
We used \tess\ short cadence and archival \galex\ observations of \textcolor{black}{1250} main sequence M stars to compare their white-light and UV flaring activity. We used these observations to test the \textcolor{black}{\galex} UV predictions of six empirical flare models that are calibrated using white-light flare observations. We did this for partially convective and fully convective M stars, in both the \galex\ NUV and FUV bandpasses. Here we present the results of our flare searches in both the \tess\ and \galex\ datasets, and then the results of our model testing. The results of our testing are shown in Tabs.\,\ref{tab:results} and \ref{tab:new_results}.

\subsection{Flares detected with \tess}
We used the method outlined in Sect.\,\ref{sec:tess_flare_detect} to search for and verfiy flares in the \tess\ 2-minute cadence lightcurves. The total number of flares detected in each sample is shown in Tab.\,\ref{tab:fitted_ffd}. 
We measured the flare rate of each sample \textcolor{black}{and the results of our fitting are} 
shown in Tab.\,\ref{tab:fitted_ffd}. We measured power law indices, $\alpha$, above 2 for all our measured samples. An $\alpha$ value above 2 indicates that small flares such as micro- and nanoflares dominate the flare energy distribution \citep[e.g.][]{Gudel03}. Along with this, it suggests that the corona of both partially and fully convective M stars are heated by successions of small flare events \citep[e.g.][]{Doyle85,Dillon20}. \textcolor{black}{Previous works have measured a range of $\alpha$ values for M stars. \citet{Medina22} measured an average $\alpha$ of $1.984\pm0.019$ for 0.1-0.3\Msun\ M stars from \tess\ photometry, \citet{Howard19} measured values between 1.84 and 2.25 for M stars from ground-based photometry and \citet{Hilton11} and \citet{Hawley14} measured values between 1.53 and 2.06. Our values sit at the top end of this range, suggesting our field age M star sample has a preference towards lower energy flares. However, as noted by \citet{Ilin21} the use of different flare detection algorithms and methods to measure $\alpha$ frustrates a direct comparison of values.}

\subsection{Flares detected with \galex}
In Sect.\,\ref{sec:galex_flare_detect_method} we outlined a method, based on the one from \citet{Brasseur19}, to search for flares in our \galex\ lightcurves. We used this method for to search for flares in both the NUV and FUV lightcurves for main sequence partially and fully convective M stars. For the partially convective stars we detected seven flares in the \textcolor{black}{\galex} NUV and one flare in the \textcolor{black}{\galex} FUV \textcolor{black}{lightcurves.} We detected 14 flares in the \textcolor{black}{\galex} NUV and four flares in the \textcolor{black}{\galex} FUV from fully convective M stars. Consequently, we were not able to use our results to test the \textcolor{black}{\galex} FUV predictions of each model for partially convective M stars. 

\subsection{Model Predictions and \textcolor{black}{UV Energy} Correction Factors} \label{sec:test_results}
As we discussed in Sect.\,\ref{sec:model_testing}, we used the \tess\ 2-min cadence and time-tagged \galex\ archival observations to test the UV predictions of six different flare models. These models and the fraction of energy each emits in the \textcolor{black}{\galex} UV \textcolor{black}{bandpasses} relative to the bolometric energy of a 9000\,K blackbody are listed in Tab.\,\ref{tab:uv_fractions}. 
For each model we used flare injection and recovery tests to simulate the predicted observed \textcolor{black}{\galex} NUV and FUV flare rate, and compared these to the average observed \textcolor{black}{\galex} UV flare rates. The results of these tests are given in Tab.\,\ref{tab:results}. The predicted and observed \textcolor{black}{\galex} NUV flare rates for partially and fully convective M stars are shown in Fig.\,\ref{fig:early_m_nuv_rates} and Fig.\,\ref{fig:mid_m_nuv_rates} respectively. The UV flare rate for the fully convective \textcolor{black}{\galex} FUV sample is shown in Fig,\,\ref{fig:mid_m_fuv_rates}.

We also followed the methods outlined in Sect.\,\ref{sec:method_correction} and ran a second round of flare injection and recovery tests with a range of increasingly energetic flare rates. 
We used these to calculate correction factors for each model, \textcolor{black}{shown in Tab.\,\ref{tab:new_results}.} These correction factors can be used to bring the predicted \textcolor{black}{\galex} UV flare rates in line with observations.

\subsubsection{9000\,K Blackbody} \label{sec:results_9000}
In Figs.\,\ref{fig:early_m_nuv_rates} and \ref{fig:mid_m_nuv_rates} we showed the results of our \galex\ injection and recovery tests for the 9000\,K blackbody in the \textcolor{black}{\galex} NUV for our partially and fully convective M star samples. We could not recover enough flares from \textcolor{black}{\galex} FUV injection tests for the fully convective sample to measure an observed discrepancy. 
We can see that for both partially and fully convective stars, this model underestimates the observed average NUV and FUV flare rates. This model underestimates the observed \galex\ NUV flare energies by a factor of $5.0^{+9.3}_{-2.5}$ for partially convective M stars, and by at least a factor of 84 for fully convective M stars. We attribute a part of this increased discrepancy to the decreased detection efficiency at lower flare energies, which result in us retrieving fewer injected events.

When we attempted to account for the changing effects of the detection efficiency on our flare injection and recovery tests, we measured best fitting \textcolor{black}{ECFs} of \textcolor{black}{$2.7\pm0.6$} and \textcolor{black}{$6.5\pm0.7$} for our partially and fully convective M star \textcolor{black}{\galex} NUV samples. 
During our analysis we noted that the observed rates appeared to have shallower slopes than our injected rates. This had the biggest effect on our fitting at the highest UV flare energies. While our fitted rate for partially convective M stars was consistent with observations at all energies, the effect was more pronounced for fully convective M stars, which were inconsistent above observed \textcolor{black}{\galex} NUV energies of \textcolor{black}{$2\times10^{31}$} ergs. This result in shown in Fig.\,\ref{fig:predicted_nuv_ecf}. 
At energies higher than this, our reported \textcolor{black}{\galex} NUV correction factors for fully convective M stars should be considered as a lower limit. 
A difference in the slope of the flare rate may indicate a change in the overall flare spectrum with energy, or the mechanism that gives rise to flares. We discuss this further in Sect.\,\ref{sec:discussion_slope}.

In the \textcolor{black}{\galex} FUV, this model does not inject enough flares at retrievable energies for us to measure a discrepancy between the predicted and observed rates. We measured an ECF of \textcolor{black}{$30.6\pm10.0$} for our fully convective M star sample. We can see this in Fig.\,\ref{fig:mid_m_fuv_rates}, but note a difference in slope such as we observed in the \textcolor{black}{\galex} NUV. The increase in \textcolor{black}{the \textcolor{black}{\galex} FUV} ECF relative to the NUV \textcolor{black}{may be} due to a combination of the contribution from emission lines, increased flare temperatures, or potential FUV flares without white-light counterparts, something we discuss further in Sect.\,\ref{sec:fuv_discussion}.

\subsubsection{Adjusted 9000\,K Blackbody} \label{sec:results_adjusted}
The adjusted blackbody model uses a 9000\,K blackbody to describe the continuum emission, but then divides the calculated energies by 0.9 to incorporate the flux from optical and UV emission lines \citep[][]{Osten15}. 
This model underestimated the observed \textcolor{black}{\galex} NUV energies of partially and fully convective M stars by factors of $4.3^{+8.7}_{-2.1}$ and \textcolor{black}{$61.2^{+175.9}_{-44.7}$} respectively. We calculated \textcolor{black}{\galex} NUV ECFs of $2.4\pm0.5$ and $5.8\pm0.6$ respectively. We did not retrieve enough flares in the \textcolor{black}{\galex} FUV to measure the difference between the predicted and observed flare rate for fully convective M stars. We calculated a \textcolor{black}{\galex} FUV ECF of \textcolor{black}{27.6$\pm$9.0} for fully convective M stars. 

This model matches the observed \galex\ UV flare rates better than the 9000\,K blackbody model, with a smaller ECF and reduced discepancy between the predicted and observed UV flare rates. We attribute this to the inclusion of flux from emission lines, alongside the 9000\,K blackbody continuum. However, it still requires ECFs of up to 5.8 in the \textcolor{black}{\galex} NUV, and a factor of $27.6\pm9.0$ in the \textcolor{black}{\galex} FUV for fully convective M stars, 
to bring its predictions in line with observations. 
One reason for this may be that while this model does include some flux from emission lines, it still underestimates the total contribution. \textcolor{black}{We discuss this further in Sect.\,\ref{sec:inferring_activity}.}

\subsubsection{9000\,K Blackbody plus the 1985 AD Leo Flare} \label{sec:results_ad_leo}
Figs.\,\ref{fig:early_m_nuv_rates} and \,\ref{fig:mid_m_nuv_rates} show the results of our tests of three AD Leo Great Flare models in the \textcolor{black}{\galex} NUV (models 3, 4 and 5). These models combine a 9000\,K blackbody with optical+UV flare spectra of the 1985 flare from AD Leo \citep[][]{Hawley91}. The blackbody and spectra are joined in the U band, with the spectra renormalised to match the estimated U band energy of the blackbody. 
The \textcolor{black}{energy} fraction used to calculate the U band emission of the blackbody varies between studies. We ran our model three times, each for a different U band emission fraction. These values were 6.7 per cent \citep[][]{Ducrot20}, 7.6 per cent \citet{Gunther20} and 11 per cent \citep[][]{Glazier20}.

\textcolor{black}{We find a range of differences between the predicted and observed NUV flare rates for our tested \textcolor{black}{sub}models.} 
For partially convective M stars we underestimate the \textcolor{black}{\galex\ NUV} energies by factors of $>120$, $>110$
and $40.0^{+51.9}_{-18.5}$ for each submodel. We do not retrieve enough flares for fully convective M stars to measure differences between the predicted and observed rates for the first two submodels. We measured a difference in energy of $>650$ for the third submodel. 
We measured ECFs of $8.2\pm1.8$, $7.3\pm1.6$ and $5.0\pm1.1$ for the partially convective M stars. We measured \textcolor{black}{\galex} NUV ECFs of $20.2\pm2.0$, $19.8\pm1.8$ and $12.4\pm1.3$ for the fully convective M stars.

In the FUV, none of these \textcolor{black}{sub}models provide enough recoverable flares to measure energy discrepancies for the partially convective M stars \textcolor{black}{or the fully convective M stars.} We measure \textcolor{black}{\galex} FUV ECFs of \textcolor{black}{$78.6\pm25.7$}, \textcolor{black}{$65.5\pm21.4$} and \textcolor{black}{$45.8\pm15.0$} for fully convective M stars. Like the previous models, the discrepancy is increased in the FUV relative to the NUV. However, we note that the difference between \textcolor{black}{\galex} FUV and NUV ECFs is not as much as for previous models. We attribute this to the use of empirical spectra to model the FUV emission, which directly includes the flux from emission lines. 

These models predict the least UV emission of all our tested models, and the results above highlight both this and the effect of the detection efficiency on our flare recovery tests. These models require larger ECFs than the 9000\,K blackbody, despite their use of empirical UV flare spectra. We attribute this to the use of a U band energy of $10^{34}$ erg \citep[][]{Segura10,Rimmer18} \textcolor{black}{instead of the original $10^{33.8}$ erg corresponding to the model spectra \citep[][]{Hawley91}} and the assumed energy fractions of the U band emission, which are used to normalise the archival flare spectra. As shown in Fig.\,\ref{fig:flare_models}, the wavelengths covered by the U band include several Balmer emission lines and the Balmer jump. Studies calculating the U band energy as a fraction of the continuum emission alone will consequently underestimate the energy in this region, in turn underestimating the UV flux from the AD Leo flare spectrum. As the position of the abiogenesis zone from \citet{Rimmer18} is dependent on the amount of NUV flux available over 2000-2800\AA, white-light flare studies using this model may underestimate the viability of prebiotic photochemistry of the surfaces of rocky exoplanets around low-mass stars. This is something we discuss further in Sect.\,\ref{sec:habitability}.

\subsubsection{MUSCLES Flare Model} \label{sec:results_muscles}
\begin{figure}
    \centering
    \includegraphics[width=\columnwidth]{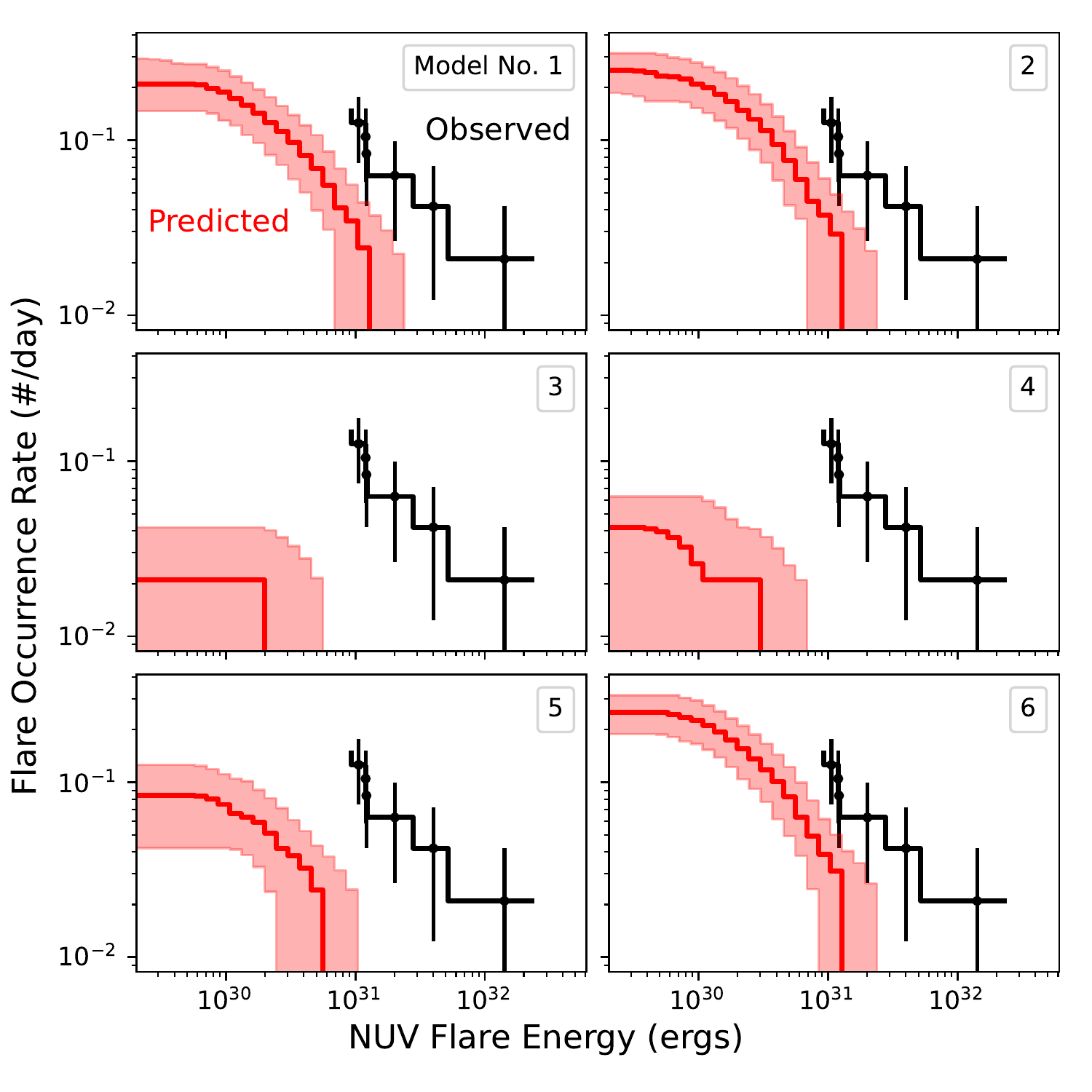}
    \caption{The results of our flare injection and recovery tests for early M stars. The black line is the observed average NUV flare rate from \galex. The red line in each plot is the predicted average NUV flare rate, obtained from our injection and recovery tests. The shaded region indicates the 1$\sigma$ uncertainties on the predicted flare rate. The number in the top right of each plot denotes the model being tested, where (1) is the 9000\,K blackbody model. Note how all models underestimate the observed NUV flare energies, with the models used for estimating the abiogenesis zone (3, 4 and 5) performing the worst.}
    \label{fig:early_m_nuv_rates}
\end{figure}

\begin{figure}
    \centering
    \includegraphics[width=\columnwidth]{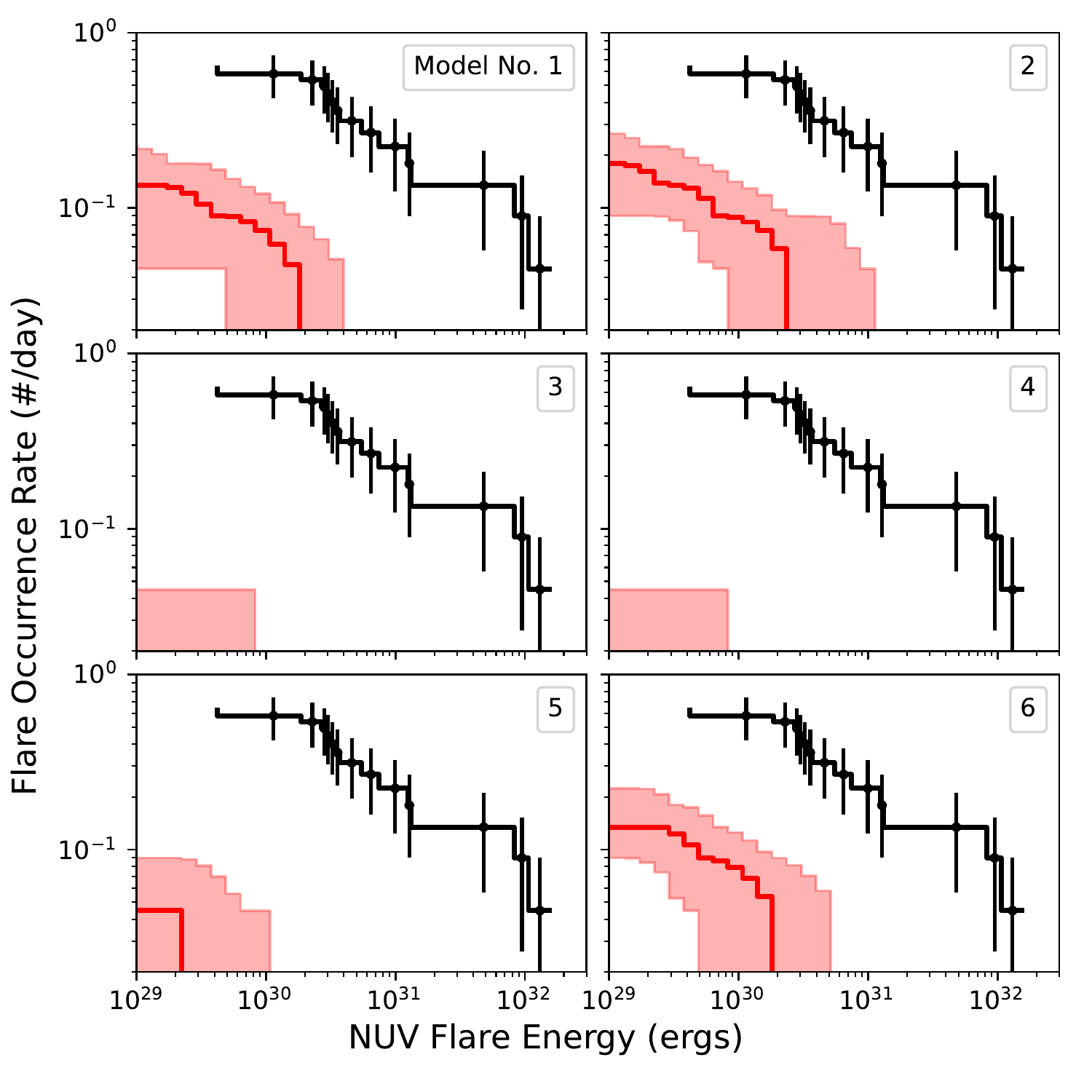}
    \caption{The results of our NUV flare injection and recovery tests for fully convective M stars. The layout is the same as in Fig.\,\ref{fig:early_m_nuv_rates}. All models underestimate the observed NUV flare energies by greater factors than for partially convective M stars.}
    \label{fig:mid_m_nuv_rates}
\end{figure}

The results of the MUSCLES flare model for each mass range in both the \textcolor{black}{\galex} NUV and FUV are denoted as model 6 in Fig.\,\ref{fig:early_m_nuv_rates} and \ref{fig:mid_m_nuv_rates}. This model uses a 9000\,K to describe the continuum emission in the optical and NUV. In addition to the blackbody continuum, the \textcolor{black}{\galex} NUV section of this model includes emission from the Mg II h\&k emission lines. The FUV section of this model was constructed using the energy budget of individual FUV emission lines, measured from  \hst\ spectra of flares from M dwarfs. It includes both the FUV flare continuum and emission lines \citep[][]{Loyd18muscles}. 

We found that this model provided the closest match to the observed average UV flare rates, but still underestimated the observed flare energies and rates. In the \textcolor{black}{\galex} NUV, this model underestimated the observed flare energies by factors of $4.1^{+6.6}_{-1.9}$ and \textcolor{black}{$81.2^{+311.2}_{-53.0}$} for partially and fully convective M stars respectively. We calculated \textcolor{black}{\galex} NUV ECFs of $2.3\pm0.5$ and $5.7\pm0.6$ for partially and fully convective M stars. Like the other models, we were unable to recover enough injected flares to measure the difference between the predicted and observed flare rate in the FUV. However, we were able to measure a \textcolor{black}{\galex} FUV ECF of \textcolor{black}{9.2$\pm$3.0}.

While this model still underestimates the observed \textcolor{black}{\galex} UV flare energies, it provides the closest match in terms of the predicted and observed \galex\ flare rates, and the required ECFs. This model provides a notable improvement in the FUV relative to the other mdoels. We attribute this to the elevated FUV continuum present in this model and the contributions from FUV emission lines. 

\begin{figure}
    \centering
    \includegraphics[width=\columnwidth]{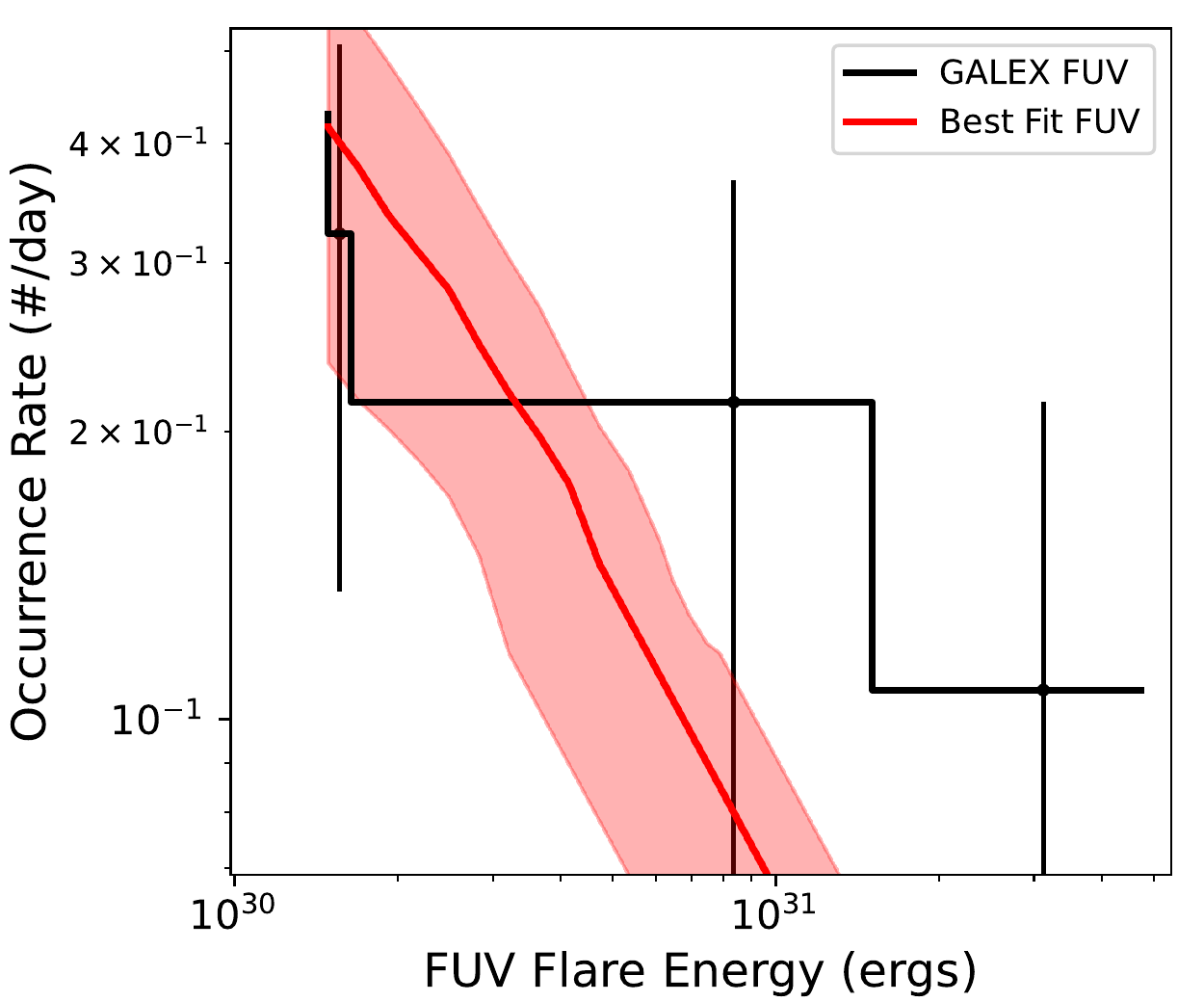}
    \caption{The best fitting FUV flare rate for fully convective stars. \textcolor{black}{This is from our second round of injection recovery tests that we used to calculate the energy correction factors. These tests are performed using UV energy fractions relative to the 9000\,K blackbody model, but do not assume a specific model}. This flare rate requires a \textcolor{black}{UV energy fraction, or } energy correction factor, of $30.6\pm10$ times that of a 9000\,K blackbody.}
    \label{fig:mid_m_fuv_rates}
\end{figure}

\begin{figure}
    \centering
    \includegraphics[width=\columnwidth]{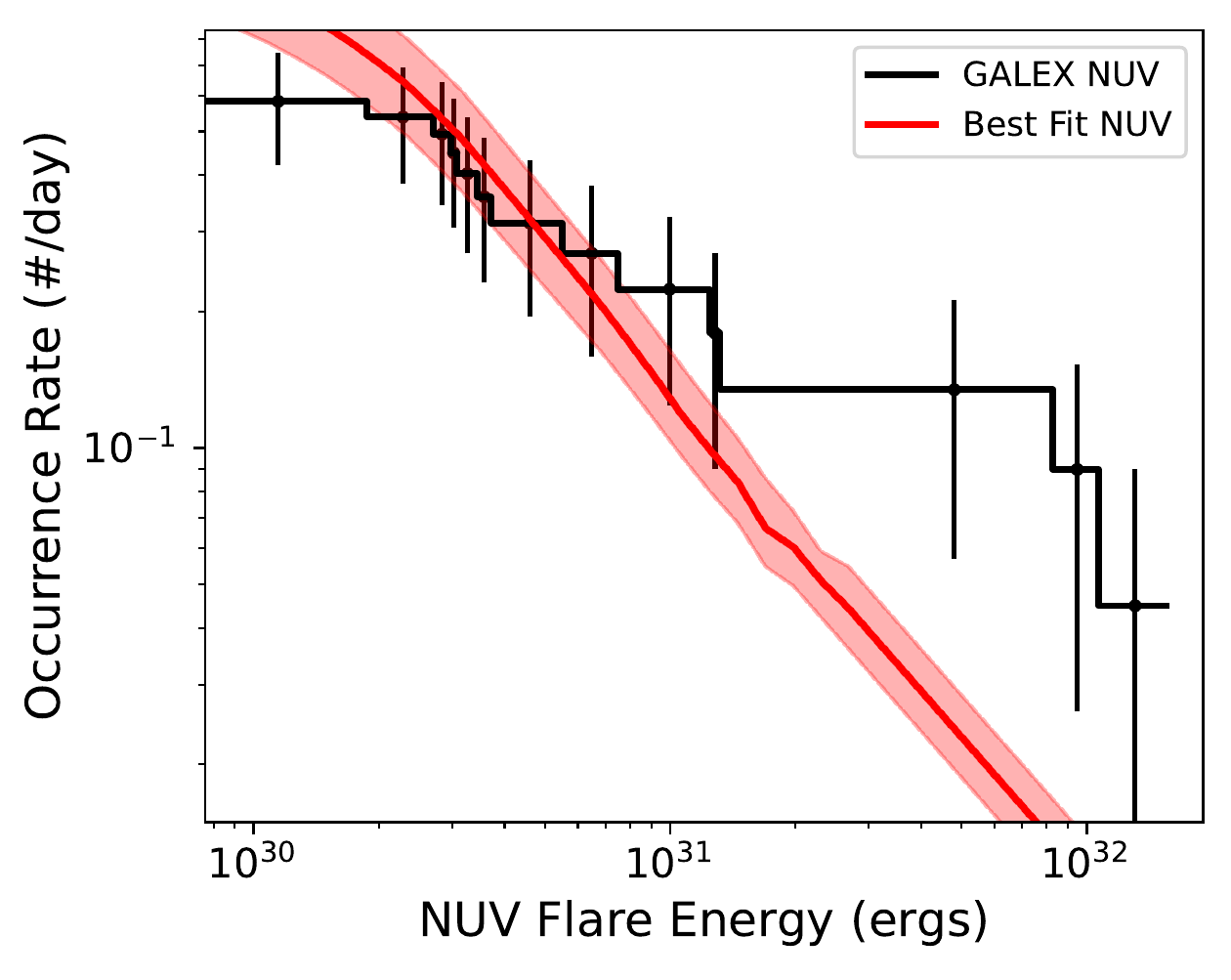}
    \caption{The best fitting predicted NUV flare rate for fully convective M stars. This uses an energy correction factor of $6.5\pm0.7$ times the NUV emission predicted by the 9000\,K blackbody. The predicted NUV flare rate (in red) shares its slope with the fitted white-light flare rate. As discussed in Sect.\,\ref{sec:results_9000}, the difference in slopes results in the predicted NUV rate underestimating the flare energies above $2\times10^{31}$ erg.}
    \label{fig:predicted_nuv_ecf}
\end{figure}

\begin{table*}
    \centering
    \begin{tabular}{c|c|c|c|}
    \hline
    Model Number & Model Name & NUV & FUV 
    \tabularnewline \hline
    M0V-M2V & &
    \tabularnewline \hline
    1 & 9000\,K Blackbody & $5.0^{+9.3}_{-2.5}$ & N/A \tabularnewline 
    2 & Adjusted blackbody & $4.3^{+8.7}_{-2.1}$ & N/A \tabularnewline
    3 & AD Leo Great Flare, 1 & $>120$ & N/A \tabularnewline
    4 & AD Leo Great Flare, 2 & $>110$ & N/A \tabularnewline
    5 & AD Leo Great Flare, 3 & $40.0^{+51.9}_{-18.5}$ & N/A \tabularnewline
    6 & MUSCLES Model & $4.1^{+7.4}_{-2.0}$ & N/A \tabularnewline
    \hline
    M4V-M5V & &
    \tabularnewline \hline
    1 & 9000\,K Blackbody & $>84$  & N/A \tabularnewline
    2 & Adjusted blackbody & $61.2^{+175.9}_{-44.7}$ & N/A \tabularnewline
    3 & AD Leo Great Flare, 1 & N/A & N/A \tabularnewline
    4 & AD Leo Great Flare, 2 & N/A & N/A \tabularnewline
    5 & AD Leo Great Flare, 3 & $>650$ & N/A \tabularnewline
    6 & MUSCLES Model & \textcolor{black}{$81.2^{+311.2}_{-53.0}$} & N/A \tabularnewline    
    \hline
    \end{tabular}
    \caption{The energy ratios between the predicted and observed flare rates calculated from our analysis. The numbers after each AD Leo Great Flare model indicate a different U band fraction used to join the UV flare spectrum and 9000\,K blackbody spectrum, and is outlined in Sect.\,\ref{sec:model_ad_leo}. 
    N/A indicates subsets where we could not recover injected flares from the \galex\ lightcurves. We attribute this to the diminished detection efficiency at lower flare energies.}
    \label{tab:results}
\end{table*}

\begin{table*}
    \centering
    \begin{tabular}{c|c|c|c|}
    \hline
    Model Number & Model Name & NUV & FUV 
    \tabularnewline \hline
    M0V-M2V & &
    \tabularnewline \hline
    1 & 9000\,K Blackbody & $2.7\pm0.6$  & N/A \tabularnewline 
    2 & Adjusted blackbody & $2.4\pm0.5$ & N/A \tabularnewline
    3 & AD Leo Great Flare, 1 & $8.2\pm1.8$ & N/A \tabularnewline
    4 & AD Leo Great Flare, 2 & $7.3\pm1.6$ & N/A\tabularnewline
    5 & AD Leo Great Flare, 3 & $5.0\pm1.1$ & N/A\tabularnewline
    6 & MUSCLES Model & $2.3\pm0.5$ & N/A \tabularnewline
    \hline
    M4V-M5V & &
    \tabularnewline \hline
    1 & 9000\,K Blackbody & $6.5\pm0.7$* & $30.6\pm10.0$ \tabularnewline
    2 & Adjusted blackbody & $5.8\pm0.6$* & $27.6\pm9.0$ \tabularnewline
    3 & AD Leo Great Flare, 1 & $20.2\pm2.0$* & $78.6\pm25.7$ \tabularnewline
    4 & AD Leo Great Flare, 2 & $19.8\pm1.8$* & $65.5\pm21.4$ \tabularnewline
    5 & AD Leo Great Flare, 3 &  $12.4\pm1.3$*  & $45.8\pm15.0$ \tabularnewline
    6 & MUSCLES Model & $5.7\pm0.6$* & $9.2\pm3.0$  \tabularnewline
    \hline
    \end{tabular}
    \caption{The energy correction factors between the predicted \textcolor{black}{\galex} UV flare rate and the best fitting UV flare rate model. This can be considered the \textcolor{black}{scaling} factor required to bring the prediction of each model in line with observations. The asterisk indicates that this is a lower limit for the correction factor, due to the different flare rate slopes in the white-light and NUV. We discuss this in detail in Sect.\,\ref{sec:discussion_slope}.}
    \label{tab:new_results}
\end{table*}

\section{Discussion}
We have presented the results of tests of the \textcolor{black}{\galex} UV predictions of empirical flare models. These models represent those used in white-light flare and habitability studies to estimate the UV \textcolor{black}{flare activity of} 
low-mass stars. We calibrated each model using white-light flare observations from \tess, and compared the predicted UV flare rates to the measured average rates from \galex\ for the same sets of stars. We found that the models used by flare and habitability studies underestimate the \textcolor{black}{\galex} UV energies and rates of flares from M stars, and we calculated ECFs that can be applied to future UV predictions to help bring them in line with observations.

\subsection{Inferring UV Flaring Activity from Optical Observations} \label{sec:inferring_activity}
In Sect.\,\ref{sec:results} we presented the results of our testing of the UV predictions of each flare model. We found that, when these models had been calibrated with white-light observations from \tess, they underestimated the NUV and FUV energies of flares observed with \galex. 
We calculated ECFs for our models, shown in Tab.\,\ref{tab:new_results}, that can be used to adjust the \textcolor{black}{\galex} UV predictions of models calibrated using \tess\ white-light flare rates. 
We found that for both the partially and fully convective M star samples, the \textcolor{black}{\galex} FUV ECFs are much larger than those for the NUV. \textcolor{black}{Flare models do a poorer job of modelling the \textcolor{black}{\galex} FUV than the NUV emission. }
We also found that the \textcolor{black}{\galex} NUV flare rates exhibited shallower slopes than those measured in the optical. This can be seen for our fully convective M star sample in Fig.\,\ref{fig:predicted_nuv_ecf}, which limited the ECF we measured to a lower limit. 

We assumed in Sect.\,\ref{sec:galex_inject} that each UV flare had a white-light counterpart and that their energies scaled linearly with each other. 
The white-light and the NUV \textcolor{black}{flare} emission are \textcolor{black}{thought to both arise from }
heated upper-photospheric and chromospheric layers \citep[][]{Fletcher07,Kowalski18}. 
These heated regions are responsible for the observed blackbody continuum emission. However, these regions also exhibit line emission, notably from Hydrogen recombination. This is responsible for the Balmer series in optical wavelengths and the consequent Balmer jump at around 3640\AA. The enhanced continuum level due to the Balmer jump has been observed at NUV wavelengths for Solar flares \citep[e.g.][]{Dominique18,Joshi21}, resulting in flux above that expected from the thermal blackbody alone. 
\citet{Kowalski19} used \hst\ NUV and ground-based optical spectroscopy of flares from the active M4 star GJ 1243 to measure the contribution of the Balmer jump to the NUV continuum. They found that a 9000\,K blackbody, when fit to the blue optical continuum, underestimated the NUV continuum by a factor of 2, and the total NUV flux in 2510-2841\AA\ by a factor of 3. The increase in the NUV continuum was attributed to the Balmer jump, and lines contributed to about 40 per cent of the observed NUV flux. \citet{Hawley07} used \hst\ NUV spectroscopy to study flares from YZ CMi. They measured line contributions ranging between 10 and 50 per cent, \textcolor{black}{but found that the most energetic flares more continuum dominated.} 
In Sect.\,\ref{sec:results_9000} we measured ECFs for the 9000\,K blackbody model of $2.7\pm0.6$ and $6.5\pm0.7$ for partially and fully convective M stars respectively. Our value for partially convective M stars is consistent with the discrepancy measured by \citet{Kowalski19}, \textcolor{black}{and greater than the 10\% line flux contribution from \citet{Osten15} used for our adjusted blackbody model.} \textcolor{black}{This suggests that for some flares it is a lack of a Balmer jump and line emission that drives the discrepancy between predictions and \textcolor{black}{\galex} NUV observations for the 9000\,K blackbody model. However, we note that for the highest energy flares that may be more continuum dominated, other sources of UV flux must be considered. } 
The \textcolor{black}{lack of a} Balmer jump and other emission lines cannot fully explain \textcolor{black}{any of our} results for fully convective M stars. This suggests the presence of extra UV flares not accounted for in our analysis and/or an extra source of UV emission not available from models. 

\subsubsection{Flare Temperatures} \label{sec:discussion_flare_teff}
Something that could explain the higher than predicted flux in the \textcolor{black}{\galex} NUV \textcolor{black}{bandpass} for \textcolor{black}{M stars} 
are flare temperatures above 9000\,K. An increase in the flare temperature will result in higher bolometric energies and a greater fraction of the continuum emission being emitted in the UV \citep[][]{Hawley07,Kowalski13}. Multi-colour and spectroscopic observations of flares from active M dwarfs have found evidence of continuum blackbody temperatures beyond 9000\,K, specifically during the impulsive and peak phase of flares, suggesting it is these regions which drive the bulk of the UV emission \citep[][]{Loyd18,Froning19,Howard20}. \citet{Froning19} measured a temperature of 40,000\,K for an $E_{FUV} = 10^{30.75}$ erg flare from GJ 674. We note that while this energy was measured over 1070-1360\AA, wavelengths shorter than those covered by the \galex\ FUV bandpass, its energy
is comparable to flares we detected in our \galex\ FUV sample (e.g. Fig.\,\ref{fig:mid_m_fuv_rates}). These measurements in both the UV and optical \citep[][]{Kowalski13,Howard20} show it is possible that our results for fully convective M stars could be explained by a flare continuum with a temperature above 9000\,K.

To test what blackbody temperature would be required to match the predicted and observed \textcolor{black}{\galex} NUV flare rates, we fitted a ``pseudo-continuum'' to the white-light and \textcolor{black}{\galex} NUV flare fluxes for both the partially and fully convective M star samples. 
The pseudo-continuum is the blackbody that is required to describe the flare emission in both the optical and the UV. As we are using \galex\ photometry, we are unable to resolve the relative contributions from emission lines and continuum in the UV. Consequently, \textcolor{black}{our fitted pseudo-continuum }
will overestimate the true blackbody flare temperature.  
To measure the best fitting pseudo-continuum flare temperature, we used the \textcolor{black}{\galex} NUV ECFs calculated for the 9000\,K blackbody model. These were $2.7\pm0.6$ and $6.5\pm0.7$ for the partially and fully convective M stars respectively. We generated a grid of \textcolor{black}{1000} blackbody curves with effective temperatures ranging between 6000 and 50,000\,K \textcolor{black}{\citep[e.g.][]{Kerr14,Howard20}}. We required each of these blackbodies to emit the same total amount of flux within the \tess\ bandpass as a 9000\,K blackbody. 
This was to make sure that a study assuming a 9000\,K blackbody flare spectrum to calculate the bolometric energy, as we did in Sect.\,\ref{sec:tess_flare_detect}, would measure the same flare power law. We then calculated the fraction of the total energy emitted in both the \galex\ NUV and FUV bandpasses for each renormalised blackbody curve. We divided the \textcolor{black}{\galex} NUV energy for each flare by the \textcolor{black}{\galex} NUV energy for a 9000\,K blackbody. \textcolor{black}{We} compared this calculated ratio to \textcolor{black}{the} \textcolor{black}{\galex} NUV ECF \textcolor{black}{for the 9000\,K blackbody model} for each mass subset. 

\begin{figure}
    \centering
    \includegraphics[width=\columnwidth]{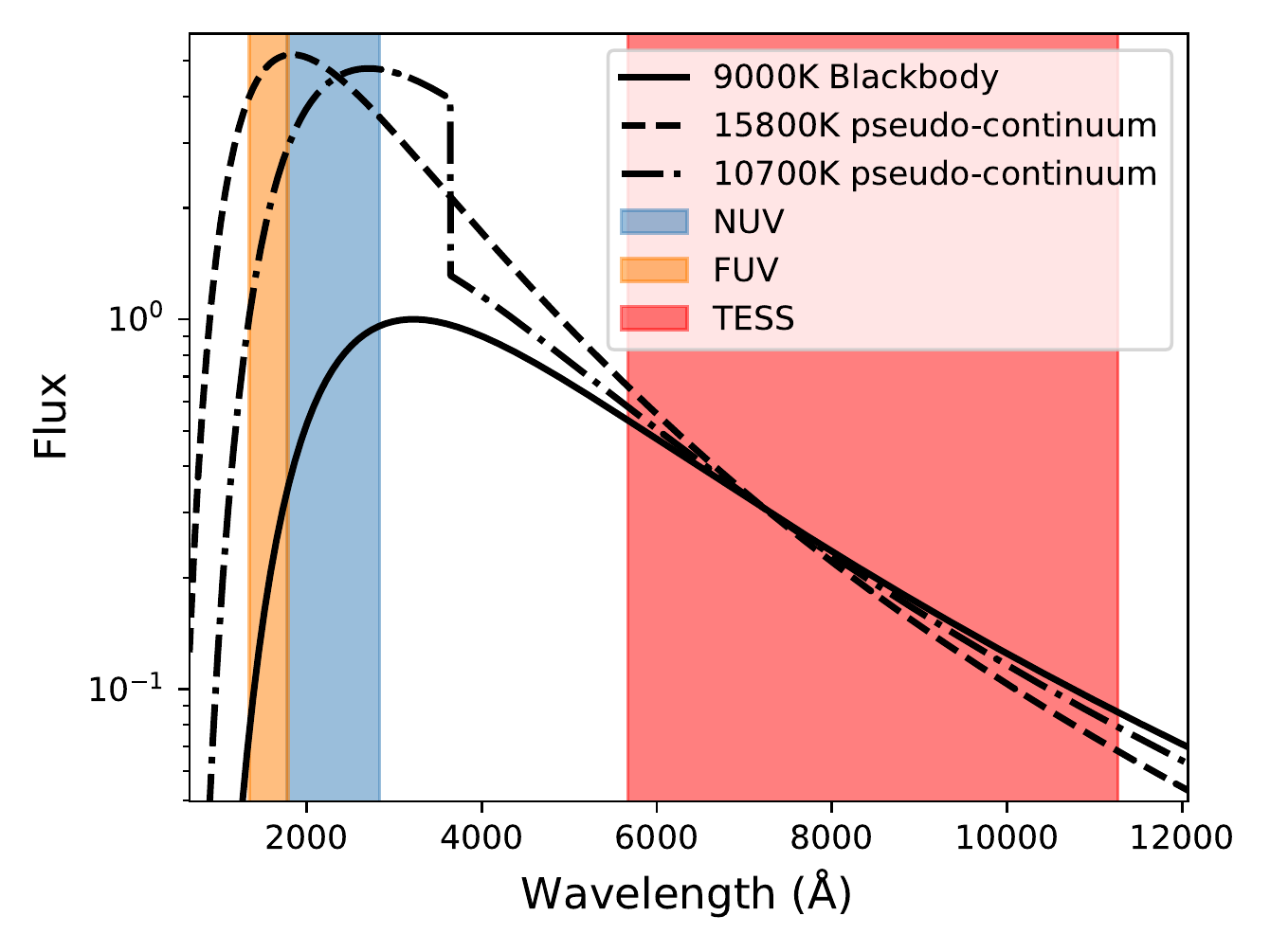}
    \caption{The best fitting pseudo-continua for fully convective M stars versus a 9000\,K blackbody. The 9000\,K blackbody has been normalised to have a maximum value of unity. We required each pseudo-continuum to give the same amount of flux in the \tess\ bandpass. As in Fig.\,\ref{fig:flare_models}, the orange, blue and red shaded areas indicates the wavelengths covered by the \galex\ FUV, NUV \textcolor{black}{and \tess\ } bandpasses. Each pseudo-continuum emits 6.5 times as much NUV flux as the 9000\,K blackbody spectrum, however the lower temperature psuedo-continuum increased by a factor of 3 at the Balmer jump to account for the flux from NUV emission features.} 
    \label{fig:pseudocont}
\end{figure}

We measured best fitting pseudo-continuum temperatures of \textcolor{black}{11,500}\,K for flares from partially convective M stars, and \textcolor{black}{15,800}\,K for fully convective M stars. 
We plotted the best fitting pseudo-continuum for fully convective M stars, alongside a 9000\,K blackbody curve, in Fig.\,\ref{fig:pseudocont}. We can see in  Fig.\,\ref{fig:pseudocont} how the increased flare temperature results in the blackbody spectrum peaking in the \galex\ UV bandpasses. 
We noted above that the pseudo-continuum fits the total flux from both the UV continuum and emission lines \textcolor{black}{and }will overestimate the true continuum temperature. To account for this, we multiplied the \textcolor{black}{\galex} NUV contribution of each \textcolor{black}{pseudo-continuum} model \textcolor{black}{used in our temperature fitting} by a factor of \textcolor{black}{three} to include the excess flux from emission lines and the elevated Balmer continuum \citep[e.g.][]{Kowalski10,Davenport12,Kowalski19}. By doing so we account for the increased NUV flare energies in our partially convective M star sample. For our fully convective M stars we measure a best fitting temperature of \textcolor{black}{10,700}\,K. The adjusted flare model can be seen in Fig.\,\ref{fig:pseudocont}. This value is greater than the typically assumed average flare temperature of 9000\,K, but consistent with the average temperatures of flares measured by previous studies \citep[e.g.][]{Howard20}. The two-component flare model shown in Fig.\,\ref{fig:pseudocont}, which attempts to include the flux from both the Balmer jump and other emission lines, is also analogous to the two-component flare model used by \citet{Kowalski10} and by \citet{Davenport12} for studying flares from YZ CMi and within SDSS stripe 82 respectively. Under the assumption that all NUV flares have a white-light counterpart, these results signify that flare studies using a single blackbody will underestimate not just the UV energy of flares, but also underestimate the bolometric energy of the thermal flare emission.

\subsubsection{Slopes of NUV Flare Rates} \label{sec:discussion_slope}
In Sect.\,\ref{sec:results_9000} we noted that the slopes of the predicted and observed \textcolor{black}{\galex} NUV flare rates appeared to differ. This limited the use of a single correction factor to a set energy range for our fully convective M star sample. 
There exist a number of reasons for why the \textcolor{black}{\galex} NUV and white-light flare rates may exhibit different slopes. The first \textcolor{black}{possibility} is a change in the flare spectrum with energy. If the relative fraction of flux emitted in the \textcolor{black}{\galex} NUV increases with the bolometric flare energy, this would result in a shallower flare rate than predicted by a constant NUV fraction. This is because larger flares would be pushed to higher energies, while the energy of smaller flares would not change greatly. \textcolor{black}{An increase in the flare continuum temperature with energy would achieve this, as the }
hotter a flare is, the greater its relative UV contribution \textcolor{black}{and subsequent UV energy} will become. 
We used the \galex\ time-tagged photon data to measure UV flare energies that were independent of any assumed model, while we calculated the white-light energies assuming models of fixed temperature. If the temperature increased relative to our assumed models, this would result in a shallower \textcolor{black}{\galex} NUV slope. To measure what pseudo-continuum flare temperatures would be required to match the \textcolor{black}{\galex} NUV emission of the highest energy events, we used the results of the flare injection and recovery tests from Sect.\,\ref{sec:method_correction}. We identified which test gave an energy matching the highest observed energy flare for the partially and fully convective M star samples. We estimated ECFs of \textcolor{black}{20} and 38 for partially and fully convective M stars respectively. When we assumed no contribution from emission lines, these ECFs corresponded to calculated temperatures of \textcolor{black}{31,000\,K} for partially convective M stars and a lower limit of 50,000\,K for fully convective M stars. When we assumed a NUV line and Balmer continuum contribution factor of 3 we measured a temperature of 16,000\,K for partially convective M stars and 22,000\,K for fully convective M stars. These values are 
similar those measured \textcolor{black}{during the rise and} peak \textcolor{black}{phases} of flares \citep[e.g.][]{Kowalski13,Muheki20, Howard20}. 
\citet{Howard20} used two-colour photometry of flares to measure peak and average temperatures of flares from low-mass stars. They noted an increase in these properties with $g'$ energy, however the relation was strongest for the peak temperature. They also noted that higher energy flares appeared to spend longer at temperatures above 14,000\,K than their low energy counterparts, due to their higher peak temperatures and longer durations. This would have the effect of raising the average temperature of higher energy flares, giving the results seen here.

We assumed \textcolor{black}{in our analysis} that the relative contributions of the blackbody continuum and NUV emission lines remain constant with increasing temperature and energy. If the contribution from line emission and the Balmer continuum change with temperature, then this may increase the required blackbody continuum temperature. \citet{Kowalski13} and \citet{Kowalski19} studied how the relative contribution of the Balmer jump changes with colour ratio, a proxy for the flare continuum temperature. They found that flares with bluer colours, i.e. higher continuum temperatures, showed smaller contributions from Balmer jumps at the peaks of their flares. Along with this, \citet{Kowalski13} measured the continuum temperature and contribution from the Balmer jump during the peak and decay phases of 13 flares from active M stars. They found that the flux from the Balmer jump, relative to the optical continuum emission, increases during the decay as the flare cools. The flux increase due to the Balmer jump was found by \citet{Kowalski19} to persist in the NUV, suggesting an increase in Hydrogen emission lines is not the cause of our observed change in flare rate. Therefore, if the average temperature of flares does increase with energy, then we might expect the contribution from Balmer emission to decrease. \citet{Hawley07} found from \hst\ NUV spectroscopy of flares from YZ CMi that higher energy flares were more continuum dominated that their lower energy counterparts. If the average flare temperature does increase with energy, as stronger reconnection events can more effectively heat the chromospheric and photospheric layers that give rise to white-light and NUV emission, then we may expect lower contributions from line emission. This would result in pseudo-continuum temperatures more similar to our initially calculated values. However, we note that to sustain a shallower slope at all NUV energies, average flare temperatures would have to increase indefinitely. 

Another explanation for the change in slope is not a change in the flare spectrum, but a change in the mechanism that drives magnetic reconnection. 
\citet{Paudel21} used \tess\ white-light and \swift\ NUV photometry to study the relative flare rates of the M4 star EV Lac. They measured \textcolor{black}{a NUV flare rate that was shallower than the white-light flare rate.} 
They attributed this to the flare rate turning over for low energy flares, which are better probed by NUV observations \citep[][]{Mullan18}. In this scenario, the footpoints of coronal magnetic loops undergo random walks driven by chromospheric flow motions. These walks result in a twisted loop which becomes unstable and undergoes a reconnection event. Large loops corresponding to flare energies above a critical energy take longer to acquire the twist required to become unstable, while lower energy flares reconnect on shorter timescales. \citet{Mullan18} posited that this critical energy depends on the magnetic scale height and thus results in lower energy flares exhibiting a shallower flare rate. If our NUV observations only probe flares below this critical energy, then we would only observe a shallower than expected flare rate.

We found in Sect.\,\ref{sec:results_9000} that the effect of the change in slope was more prevalent for our fully convective M star sample. 
\citet{Mullan18} found that the lower electrical conductivity of late-M stars means that diffusion of magnetic field lines would also contribute to driving loop instabilities. This would have the effect of decreasing the timescale further and causing even shallower flare rates. \citet{Mullan18} noted that these effects likely set in at a spectral type of around M5, making it possible that some of the stars in our fully convective sample are affected by this, driving the increased difference between the white-light and NUV flare slopes. Alternatively, the increased magnetic field strength of fully convective M stars is able to drive more intense heating even for lower energy flare events. 
To differentiate between these scenarios, a temperature-energy relation in the UV and optical needs to be confirmed through simultaneous observations.

\subsection{Matching the FUV emission to the NUV and white-light - more energy, or more flares?} \label{sec:fuv_discussion}
In Sect.\,\ref{sec:results} we presented the \textcolor{black}{\galex} FUV results for each model. We found that the \textcolor{black}{\galex} FUV models exhibit greater differences between the predicted and observed flare rates \textcolor{black}{than in the \textcolor{black}{\galex} NUV}, with models predicting no flares at recoverable energies. We also calculated ECFs for each model, shown in Tab.\,\ref{tab:new_results}. Both \textcolor{black}{\galex} FUV samples required higher ECFs than their NUV counterparts, with the 9000\,K blackbody model requiring a minimum ECF of \textcolor{black}{$30.6\pm10.0$} for the fully convective M star sample.  Our best fitting model was the MUSCLES flare model, however this still required an ECF of \textcolor{black}{$9.2\pm3.0$} to bring its predictions in line with observations. These factors suggest that, if we assume every UV flare has a white-light counterpart, that current models require an extra source of FUV emission. 

\subsubsection{A Single High Temperature Blackbody?}
In Sect.\,\ref{sec:discussion_flare_teff} we calculated best fitting pseudo-continuum temperatures for the white-light and \textcolor{black}{\galex} NUV flare rates. We measured a best fitting pseudo-continuum temperature of \textcolor{black}{15,800\,K} for fully convective M stars. When we accounted for the contribution of the NUV Balmer continuum and NUV line emission, the required blackbody temperature decreased to \textcolor{black}{10,700\,K}. 
To test whether a single pseudo-continuum could be used to model both the missing \textcolor{black}{\galex} NUV and FUV emission we calculated the \textcolor{black}{\galex} FUV fluxes of our best fitting pseudo-continuum blackbody curves. When we do not account for the NUV line emission in flares from fully convective M stars \textcolor{black}{and use a temperature of 15,800\,K} we measure a \textcolor{black}{\galex} FUV flux of 24 times that from the 9000\,K blackbody. 
When we use the temperature of 10,700\,K and include the contribution of the elevated \textcolor{black}{Balmer} continuum to FUV emission, we calculate an \textcolor{black}{\galex} FUV flux 8.8 times that from the 9000\,K blackbody. To account for the contribution of the elevated continuum to the FUV flux, we multiply the predicted pseudo-continuum flux by 2 \citep[][]{Kowalski19}. We note here that this model only includes the contribution from an elevated continuum to the FUV flux, and does not include any FUV flux from emission lines. If we use the maximum temperature of $>50,000$\,K we calculated in Sect.\,\ref{sec:discussion_slope} for fully convective M stars, we measure an \textcolor{black}{\galex} FUV flux 
above 308 times that from a 9000\,K blackbody. 
When accounting for NUV Balmer and line emission, and including the contribution of the elevated continuum to the FUV flux, the predicted \textcolor{black}{\galex} FUV emission decreases to \textcolor{black}{150} times the \textcolor{black}{\galex} FUV flux from a 9000\,K blackbody. Our second round of injection and recovery tests were not enough to match the highest energy FUV flare. We reran this analysis for our fully convective M star \textcolor{black}{\galex} FUV sample until we obtained a flare rate that intersected with the highest energy flare. This resulted in an ECF of \textcolor{black}{200}, greater than that available from the thermal blackbody and Balmer jump continuum.

One potential explanation for the observed discrepancy between the white-light, \textcolor{black}{\galex} NUV and FUV observations is that models are underestimating the flux from FUV line and enhanced continuum emission. FUV observations of flares with \hst\ have shown that the continuum dominates the observed emission. \citet{Hawley03} measured line-to-continuum energy ratios of between 0.02 and 0.08 for flares from AD Leo. We note that they used spectra obtained with \hst\ STIS with wavelengths of 1160--1700\AA, which encompasses the \galex\ FUV bandpass but also includes regions where the blue flare continuum will emit more strongly. We are unable to match the required ECFs if we assume the lines contribute an energy budget fraction of 0.08. These observations also included the contribution from any elevated Balmer continuum. 
Solar flare observations have shown that the elevated Balmer continuum persists down to 2000\AA\ \citep[][]{Dominique18}, and simulated flare spectra show it should persist into the \galex\ FUV bandpass \citep[][]{Kowalski18}. 
As mentioned above we used the NUV continuum jump of 2 measured by \citet{Kowalski19} to calculate the above correction factors and assumed it persists into the \galex\ FUV bandpass. As noted in Sect.\,\ref{sec:discussion_slope}, the contribution of lines and the Balmer jump likely decrease with flare temperature. Therefore, we do not expect a combined contribution from line and hotter continuum temperatures is able to solve our discrepancy at any of our measured flare energies. 

\subsubsection{A Need For A New Type Of Flare Model?}

Another explanation \textcolor{black}{is that} 
the FUV and white-light emission are dominated by different parts of the flare process, and should not be considered as one single spectrum as current models assume. 
Recent simultaneous \hst\ FUV and \tess\ optical observations of a flare from Proxima Centauri showed an apparent delay between the FUV and optical emission observed with \tess\ \cite[][]{MacGregor21}. This flare had an FUV energy of $10^{30.3}$ erg and an inferred blackbody temperature from the FUV pseudo-continuum of 15,000--22,000\,K. \citet{MacGregor21} found that this was unable to explain the amplitude of the flare in the \tess\ bandpass. 
They also noted that the FUV instead was better correlated with the millimetre wavelength emission typically associated with the accelerated non-thermal electron beams. This impulsive phase emission was attributed by \citet{MacGregor21} to a hot blackbody with contribution from gyrosynchrotron emission, with the delayed white-light emission coming from the thermally heated flare footpoints. \citet{Froning19} attributed 40,000\,K FUV emission of a flare from GJ 674 to the presence of a chromospheric condensation, \textcolor{black}{a downflowing compression created during flares by jets of non-thermal electrons \citep[e.g.][]{Graham20}. The bulk of non-thermal electrons in these jets are responsible for heating this layer, while only the highest energy particles are able to penetrate and heat lower chromospheric layers. The dense chromospheric condensation starts at high temperatures, but cools as it descends until it reaches around 10,000\,K \citep[e.g.][]{Kowalski18}.} In order for their modelling to simultaneously provide a $>25,000$\,K FUV temperature and $\approx$10,000\,K optical temperature they required an extra chromospheric emitter not explained by previous radiative-hydrodynamic (RHD) models. However, we note that \citet{Froning19} did not have simultaneous FUV and optical temperature measurements, and that \citet{Howard20} measured white-light flare continuum temperatures above 20,000\,K, suggesting that the second emitter may not be strictly required. 
Regardless, these observations point towards the impulsive phase, and bulk, of the FUV emission coming from an intensely heated region at the start of the flare. This region then cools, transferring its energy to lower atmospheric layers. 
The \textcolor{black}{lower layers are heated and emit the} bulk of the observed NUV and white-light emission, along with some additional FUV flux. \textcolor{black}{A chromospheric condensation may satisfy these requirements.} 
If the FUV emission is dominated by an initial high temperature layer \citep[e.g.][]{Froning19}, and the white-light/NUV emission is dominated by a cooler region that increases in size as the condensation cools, then we might not expect a single temperature flare model calibrated to white-light data to match the FUV observations. 

One way around this problem could be to create a model that has a changing continuum temperature as part of its nature. However, the timescale of any temperature changes is likely to depend on individual flare morphologies and thus change between events \citep[e.g.][]{Kowalski13,Howard20}. Along with this, the strength of any white-light component may also change between flares of similar FUV energies. \citet{Loyd20} showed that the optically quiet M dwarf GJ 887, which showed no flares in its \tess\ lightcurve, flared twice in 2.8 hours of \hst\ FUV observations, with energies above $10^{30}$ erg. This energy is similar to that of the flare detected from Proxima Centauri by \citet{MacGregor21}, who observed a slightly delayed white-light counterpart. As GJ 887 ($T$=5.5) is brighter than Proxima Centauri ($T$=7.4), we might expect FUV flares of this energy to also have a white-light counterpart that is detectable with \tess. This suggests that while the FUV emission may be commonly detected from the initial heating, this does not always result in detectable white-light emission for flares from low-mass stars.  
Solar flares without white-light counterparts have frequently been detected in hard X-rays, which probes the emission from the non-thermal electron jets. Works comparing flares with and without white-light counterparts have found that flares with white-light counterparts tend to have stronger magnetic fields and higher energy deposition rates than those without \citep[e.g.][]{Watanabe17, Song18a}. However, \citet{Watanabe20} also found that intermediate energy non-thermal electrons may reach lower layers for direct heating if the density of the plasma in the chromosphere has been reduced, for example by a previous chromospheric evaporation. If this the case for M dwarfs, then we may only expect detectable white-light flares either from events with rapid energy deposition or a reduced local chromospheric plasma density. FUV flares then would occur whenever intermediate energy electrons generate a chromospheric condensation, but associated white-light emission will only reach detectable levels in specific cases. 
This fits with studies suggesting that most Solar flares do exhibit white-light emission, albeit at low levels \citep[e.g.][]{Hudson06,Song18b}. Depending on the temperature of the initially heated condensation, there may also be cases of NUV flares without white-light counterparts also. In this situation, the increased \textcolor{black}{\galex} NUV ECF we measured for fully convective M stars in Sect\,\ref{sec:results_9000} and used to calculate a pseudo-continuum temperature of 10,700\,K in Sect.\,\ref{sec:discussion_flare_teff} may instead point towards an excess of NUV flares. If this is the case, it would further complicate efforts to extrapolate white-light flare rates, measured from flares with various impulsivities and energies, to the FUV. Yet, how the white-light and UV energetics of M dwarf flares scale with each other, and what fraction of white-light flares have UV counterparts is unknown. Simultaneous observations of multiple flares in the optical and UV are required to answer these questions and probe the mechanisms at work during high energy flares.

\subsection{Impact on Exoplanet Habitability} \label{sec:habitability}
Our results will have an impact on the results of studies that seek to test whether prebiotic photochemistry could be viable on the surfaces of rocky exoplanets around low-mass stars. 
Due to their cool temperatures M stars may not be able to provide the NUV flux required for this photochemistry from their quiescent state alone \citep[e.g.][]{Ranjan17,Rimmer18}. These studies suggested that flares, if they occurred often enough, could potentially provide the missing flux required for this processes to occur. \textcolor{black}{\citet{Rimmer18} used spectra from the 1985 AD Leo Great Flare to measure a relation between the U band energy and NUV photon flux of flares. They combined this relation with white-light flare rates measured from \kepler\ data to determine that only the most active M stars flared enough to provide the required NUV flux for prebiotic photochemistry to be viable. }
This was later confirmed by studies who combined the U band and NUV flux relation from \citet{Rimmer18} with white-light flare rates measured from \tess, \ktwo\ and ground-based observations \citep[e.g.][]{Gunther20,Glazier20,Ducrot20}. These studies also found that a \textcolor{black}{minority} of low-mass stars exhibit flare rates high enough for flare-assisted abiogenesis. 

\citet{Gunther20} used \tess\ 2-minute cadence observations from sectors 1 and 2 to measure the flare rates of individual stars. In their sample of 401 early-M and 271 mid to late-M stars, 3 early-M and 8 mid to late-M stars had flare rates that intersected with the abiogenesis zone. 
\citet{Glazier20} and \citet{Ducrot20} used \ktwo, EvryScope and TRAPPIST observations respectively to study the flare rate of TRAPPIST-1, and found this star did not flare enough for flare-assisted abiogenesis to be viable. To calculate the U band energy for use with the relation from \citet{Rimmer18}, these studies calculated the bolometric energy by assuming a 9000\,K blackbody for the flare spectrum, and then calculating the U band energy as a fraction of this. The assumed fractions for these studies by 6.7 per cent \citep[][]{Ducrot20}, 7.6 per cent \citep[][]{Gunther20} and 11 per cent \citep[][]{Glazier20}. We found in Sect.\,\ref{sec:results_ad_leo} that 
the resultant predicted \textcolor{black}{\galex} NUV flux significantly underestimates the observed values. We attributed this to the initial assumption of a U band energy of $10^{34}$ ergs (instead of the original model value that corresponds to the UV fluxes) and the use of the U band energy to combine the 9000\,K blackbody with the AD Leo flare spectrum. 

We applied our results to the models used by these studies. By multiplying the calculated U band energies used with the \citet{Rimmer18} relations by our calculated \textcolor{black}{\galex} ECFs, we can bring the predicted NUV flux in line with observations. This reduces the U band energies required for a flare rate to overlap with the abiogenesis zone, where flare-assisted prebiotic photochemistry is viable. We did this first for the M star flare rates measured by \citet{Gunther20}. After correcting for the missing NUV emission, \textcolor{black}{16} early-M stars and \textcolor{black}{35} mid to late-M stars had flare rates that intersected with the abiogenesis zone. This increased the incidence rate for flaring early-M stars from 0.7 per cent to 4.0 per cent, and from 3.0 per cent to 13 per cent for flaring mid to late-M stars. 
These results suggest that M stars, particularly fully convective ones, do not have to be as active as previously expected in order to provide the NUV flux required for prebiotic photochemistry to be viable. \textcolor{black}{Our results in Sect.\,\ref{sec:results} show that any increase in the NUV emission will be accompanied by an increase in FUV flux. FUV photons can dissociate biosignatures and alter atmospheric compositions \citep[][]{Venot16,Rimmer19}, highlighting how all factors need to be considered when assessing exoplanet habitability.}

We also applied our results for fully convective M stars to the abiogenesis zone for TRAPPIST-1 from both \citet{Ducrot20} and \citet{Glazier20}. We found that after applying our NUV correction factors for fully convective M stars, the flare rate of TRAPPIST-1 intersected with the abiogenesis zone at bolometric flare energies of $10^{36}-10^{37}$ ergs. The flare rates measured by \citet{Glazier20} and \citet{Ducrot20} predict a $10^{36}$ erg flare from TRAPPIST-1 should occur once every 10-20 years. \textcolor{black}{A flare of this energy has not yet been observed from TRAPPIST-1. }
The highest energy \textcolor{black}{observed} flare is on the order of $10^{33}$ ergs, well below the required value for viable abiogenesis. To test whether TRAPPIST-1 could potentially exhibit a $10^{36}-10^{37}$ erg flare, we used the ASAS-SN catalogs of flares detected from mid to late-M stars from \citet{Schmidt19} and \citet{Rodriguez20}. These studies used the wide-field ASAS-SN transient survey data to search for rare high energy flares from M stars, with \citet{Schmidt19} providing a particular focus on mid to late-M stars. These studies measured V band flare energies up to $10^{35}$ erg for early to mid-M stars, and $10^{34}$ erg for late-M stars similar to TRAPPIST-1. This corresponds to a bolometric flare energy of $10^{36}$ erg for late-M stars, after using the V band to bolometric energy conversion factor from \citet{Rodriguez20}. This suggests that while rare, flares of this energy may still be possible from TRAPPIST-1. However, such a flare would be accompanied by damaging FUV irradiation. Along with this, \citet{Ilin21} showed that for fast rotating (P< 9 hours) ultracool dwarfs, giant flares occurred closer to the poles than the equator, attenuating the amount of high energy flux received by orbiting planets. If this result holds for TRAPPIST-1, then even higher energy and thus rarer flares would be needed to provide the NUV flux required for prebiotic photochemistry to occur. This highlights the importance of considering both the full flare spectrum and where the flares originate when trying to understand the potential habitability of terestrial planets around low-mass stars.

\section{Conclusions}
We have presented the results of testing the NUV and FUV predictions of flare models calibrated using white-light observations. 
Previous studies have used white-light observations of stellar flares to calibrate empirical flare models, which can then be used to predict the UV energies and rates of flares from low-mass stars. We used \tess\ observations to measure the average white-light flare rates of \textcolor{black}{1250 field} partially and fully convective M stars \textcolor{black}{that had previously been observed with \galex.} We combined the measured average flare rate with six different empirical flare models to predict the average \galex\ NUV and FUV flare rates of each sample. We measured the average NUV and FUV flare rates of each sample using archival \galex\ lightcurves generated using the gPhoton python package. 
We compared our predictions to the observed average UV flare rates of each mass subset, using flare injection and recovery techniques to account for the short visit durations and noise properties of the \galex\ lightcurves.

We tested six empirical flare models in this work. The first two were a 9000\,K blackbody spectrum and a blackbody spectrum adjusted to include a flux contribution from emission lines. The next three were based on a combination of a 9000\,K blackbody spectrum and UV spectra of the 1985 Great Flare from AD Leo.  This model combines the blackbody curve and spectra in the $U$ band, assuming a certain fraction of the blackbody emission in this wavelength region. We tested three separate models, each using a different $U$ flux fraction found in the literature. The final model was the MUSCLES model from \citet{Loyd18muscles}, which combines a 9000\,K blackbody with NUV emission lines and an FUV flare model based on \hst\ spectroscopic observations. We found that the MUSCLES model best estimates the NUV and FUV flare energies in each mass subset, but still underestimates the energies of the smaller flares in our sample by up to \textcolor{black}{$5.7\pm0.6$} and \textcolor{black}{$9.2\pm3.0$} respectively. 

We used the results of our flare injection and recovery to calculate NUV and FUV ECFs for each tested model. We used our calculated ECFs to estimate average flare temperatures. We found that the \textcolor{black}{\galex} NUV flare emission of partially convective M stars can be well explained by a 9000\,K blackbody with a contribution from the Balmer jump, while flares from fully convective M stars require a temperature of 10,700\,K and a contribution from the Balmer jump. However, we also found that the slope of the predicted flare rate was steeper than the observed rate. This effect was more pronounced for our fully convective M star sample, making our calculated correction factor a lower limit after an \textcolor{black}{\galex} NUV energy of $2\times10^{31}$ergs. We discussed what may cause this difference in the slope between the predicted and observed UV flare rates. We attributed it to either an increase of flare temperature with energy, resulting in temperatures up to 22,000\,K for our samples, or a change in the reconnection mechanism between low and high energy flares. 

The correction factors and temperatures inferred from our \textcolor{black}{\galex} NUV observations were not enough to match the observed \textcolor{black}{\galex} FUV flare energies and rate for fully convective stars. We discussed potential causes for this and attributed it to the FUV and white-light emission being dominated by different parts of the flare process, and some FUV flares lacking white-light counterparts. Simultaneous optical and FUV flare observations are required to test this connection further. 

We applied our results to tests of habitability made by previous studies. We found that studies using a combination of the 9000\,K blackbody and the AD Leo Great Flare spectrum underestimated the NUV flux of M stars by between factors of 5 and 20. When we applied our correction factors to the results from \citet{Gunther20}, we found that 4 and 13 per cent of flaring partially and fully convective M stars now had flare rates that intersected the abiogenesis zone, an increase from 0.7 and 3.0 per cent. We also applied our NUV correction factor for fully convective M stars to the flare rate calculated by \citet{Glazier20} and \citet{Ducrot20} for TRAPPIST-1. We found that TRAPPIST-1 would need to flare with an energy of $10^{36}-10^{37}$ erg to intersect with the abiogenesis zone. 
Our results highlight the need for further joint analysis of the optical and UV properties of stellar flares.

\section*{Acknowledgements}
The authors would like to thank A. Segura for sharing their flare models. \textcolor{black}{The authors also thank the referee for helping to improve this manuscript.} This research has made use of the SVO Filter Profile Service (http://svo2.cab.inta-csic.es/theory/fps/) supported from the Spanish MINECO through grant AYA2017-84089. JAGJ and ES acknowledge support from grant HST-GO-15955.004-A from the Space Telescope Science Institute, which is operated by the Association of Universities for Research in Astronomy, Inc., under NASA contract NAS 5-26555. JAGJ acknowledges support from grant HST-AR-16617.001-A from the Space Telescope Science Institute, which is operated by the Association of Universities for Research in Astronomy, Inc., under NASA contract NAS 5-26555. 
TRY would like to acknowledge additional support from the Future Investigators in NASA Earth and Space Exploration (FINESST) award 19-ASTRO20-0081. \textcolor{black}{CM acknowledges support from NASA grant 80NSSC18K0084.}

This paper includes data collected by the \tess\ and \galex\ missions, which are publicly available from the Mikulski Archive for Space Telescopes (MAST). Funding for the TESS mission is provided by NASA's Science Mission directorate. This research was supported by NASA under grant number \textcolor{black}{80NSSC22K0125} from the TESS Cycle 4 Guest Investigator Program.

\section*{Data Availability}
All the data used in this paper is publicly available through MAST.


\bibliographystyle{mnras}
\bibliography{reference} 




\bsp	
\label{lastpage}
\end{document}